\documentclass{article}
\usepackage{amsmath,amssymb, latexsym}
\usepackage[dvips]{epsfig}
\usepackage{subfigure}
\usepackage{wrapfig, psboxit}
\usepackage{verbatim}
\usepackage{epsf,mathbbol}
\usepackage{rotating}
\usepackage{ifthen}
\usepackage[usenames]{pstcol}
\usepackage{pst-node}
\usepackage{pst-plot}
\usepackage{pst-coil}
\usepackage{multido}
\usepackage{pst-3d}
\usepackage{color}
\usepackage{calc}
\newcommand{\SQRTwo}{0.717}

\newcommand{\MediumScale}{0.6}
\newcommand{\SmallScale}{0.5}

\newlength{\MediumStateDiameter}
\newlength{\SmallStateDiameter}
\newlength{\LargeStateDiameter}
\newlength{\VerySmallStateDiameter}
\newlength{\StateLineWidth}        
\newcommand{\StateLineStyle}{solid} 
\newcommand{\StateLineColor}{black}
\newif\ifStateLineDbl \StateLineDblfalse 
\newcommand{\StateLineDblCoef}{0.6} 
\newcommand{\StateLineDblSep}{0.4} 
\newcommand{\VSStateLineCoef}{.6} 
\newcommand{\StateFillStatus}{solid} 
\newcommand{\StateFillColor}{white}
\newcommand{\StateLabelColor}{black}
\newcommand{\StateLabelScale}{1.7}
\newcommand{\DimStateLineStyle}{solid} 
\newcommand{\DimStateLineCoef}{1} %
\newcommand{\DimStateLineColor}{gray}
\newcommand{\DimStateLabelColor}{gray}
\newcommand{\DimStateFillColor}{white}
\newlength{\EdgeLineWidth}
\newcommand{\EdgeLineStyle}{solid}
\newif\ifEdgeLineDbl \EdgeLineDblfalse 
\newcommand{\EdgeLineDblCoef}{0.5} 
\newcommand{\EdgeLineDblSep}{0.6} 
\newcommand{\EdgeLineColor}{black}
\newlength{\EdgeArrowWidth}
\newlength{\EdgeDblArrowWidth}
\newcommand{\EdgeArrowLengthCoef}{1.4}
\newcommand{\EdgeDblArrowLengthCoef}{1.7}
\newcommand{\EdgeArrowInset}{0.1}
\newcommand{\EdgeArrowStyle}{->}
\newcommand{\EdgeRevArrowStyle}{<-}
\newcommand{\EdgeLineBorderCoef}{2}
\newcommand{\EdgeLineBorderColor}{white}
\newcommand{\EdgeLabelColor}{black}
\newcommand{\EdgeLabelScale}{1.7}
\newcommand{\DimEdgeLineCoef}{1.2} 
\newcommand{\DimEdgeLineStyle}{solid} 
\newcommand{\DimEdgeLineColor}{gray}
\newcommand{\DimEdgeLabelColor}{gray}
\newlength{\EdgeOffset}
\newcommand{\VaucArcAngle}{15}
\newcommand{\VaucArcCurvature}{0.8}
\newlength{\VaucArcOffset}
\newcommand{\VaucLArcAngle}{30}
\newcommand{\VaucLArcCurvature}{0.8}
\newlength{\LoopOffset}
\newlength{\LoopVarOffset}
\newcommand{\LoopAngle}{30}
\newcommand{\CLoopAngle}{22}
\newcommand{\LoopVarAngle}{28}
\newcommand{\LoopOnMediumState}{7}
\newcommand{\LoopOnSmallState}{9.6} 
\newcommand{\LoopOnLargeState}{5.8}

\newcommand{\CLoopOnMediumState}{8}
\newcommand{\CLoopOnSmallState}{12}
\newcommand{\CLoopOnLargeState}{6}

\newlength{\TransLabelSep}
\newcommand{\EdgeLabelPosit}{.45}
\newcommand{\ArcLabelPosit}{.4}
\newcommand{\LArcLabelPosit}{.4}
\newcommand{\LoopLabelPosit}{.25}
\newcommand{\CLoopLabelPosit}{.25}
\newcommand{\InitStateLabelPosit}{.1}
\newcommand{\FinalStateLabelPosit}{.9}
\newcommand{\ArrowOnStateCoef}{}
\newcommand{\ArrowOnMediumState}{1.5}
\newcommand{\ArrowOnSmallState}{1.7} 
\newcommand{\ArrowOnLargeState}{1.3}
\newcommand{\ArrowOnVerySmallState}{5} 
\newlength{\VertShiftH} 
\newlength{\VertShiftD} 
\newlength{\VertShift}
\newif\ifVCFrame

\newif\ifVCGrid

\newif\ifVCRigidLabel

\newif\ifVCStateLabelBaseLine

\psset{unit=1cm}
\newpsstyle{VaucFrameStyle}{arrows=-,framesep=0pt,%
                     linewidth=0.6pt,linecolor=black,%
                                         linestyle=solid,%
                                         doubleline=false,%
                                         fillcolor=white,fillstyle=none,%
                                         cornersize=relative,framearc=0}
\newcommand{\FrameStyle}{\psset{style=VaucFrameStyle}}
\newpsstyle{VaucGridStyle}{%
      gridwidth=0.6pt,griddots=10,subgriddiv=1,%
          gridlabels=7pt}
\newcommand{\GridStyle}{\psset{style=VaucGridStyle}}
\newenvironment{VCPicture}[2][.5]%
  {\settoheight{\VertShiftH}{$\{$}%
   \settodepth{\VertShiftD}{$\{$}%
   \setlength{\VertShift}{.5\VertShiftD-.5\VertShiftH}%
   \begin{pspicture}[#1]#2%
   \ifVCFrame \FrameStyle \psframe#2\fi%
   \ifVCGrid \FrameStyle\GridStyle \psgrid#2\fi}%
  {\RstState\RstEdge%
   \end{pspicture}}
\newcommand{\VCScale}{}
\newcommand{\FixVCScale}[1]{\renewcommand{\VCScale}{#1}}

\newcommand{\MediumPicture}{\FixVCScale{\MediumScale}}
\newcommand{\SmallPicture}{\FixVCScale{\SmallScale}}

\newcommand{\VCDirectory}{}
\newcommand{\SetVCDirectory}[1]{\renewcommand{\VCDirectory}{#1}}
\newif\ifVCName

\newcommand{\VCDraw}[2][\VCGridScale]{%
\psset{unit=#1cm}%
\ifVCName\makebox[0pt][r]{\fbox{{\scriptsize #2}}}\fi%
\scalebox{\VCScale}{#2}%
\psset{unit=1cm}}

\newcommand{\VCPut}[3][0]{\rput{#1}#2{#3}}%
\newlength{\StateLineWid}
\newcommand{\StateLineSty}{\StateLineStyle} 
\newcommand{\StateLineCol}{\StateLineColor}

\newcommand{\StateFillCol}{\StateFillColor}
\newcommand{\StateFillSta}{\StateFillStatus} 
\newcommand{\StateLabelSca}{1}
\newcommand{\StateLabelCol}{\StateLabelColor}
\newcommand{\StateDimen}{outer}
\newcommand{\StateDblDimen}{middle}
\newcommand{\VCIFflag}{}
\newcommand{\PlainState}%
  {\renewcommand{\VCIFflag}{0}}
\newcommand{\FullState}%
  {\renewcommand{\VCIFflag}{2}}

\newif\ifVCShowState

\newcommand{\ShowState}{\VCShowStatetrue}
\ShowState 
\newpsstyle{VaucStateStyle}{framesep=0pt,%
         linewidth=\StateLineWid,linecolor=\StateLineCol,%
         linestyle=\StateLineSty,doubleline=false,%
         fillcolor=\StateFillCol,fillstyle=\StateFillSta,%
         border=0pt,dimen=\StateDimen,%
         cornersize=relative,framearc=1,framesep=0pt}
\newpsstyle{VaucStateDblStyle}{framesep=0pt,%
         linewidth=\StateLineDblCoef\StateLineWid,linecolor=\StateLineCol,%
         linestyle=\StateLineSty,doubleline=true,doublesep=\StateLineDblSep\StateLineWid,%
         fillcolor=\StateFillCol,fillstyle=\StateFillSta,%
         border=0pt,dimen=\StateDblDimen,%
         cornersize=relative,framearc=1,framesep=0pt}
\newpsstyle{VaucHiddenStateStyle}{framesep=0pt,%
         linewidth=\StateLineWid,linecolor=\StateLineCol,%
         linestyle=none,%
         fillcolor=\StateFillCol,fillstyle=none,%
         border=0pt,dimen=outer,%
         cornersize=relative,framearc=1,framesep=0pt}
\newcommand{\StateStyle}{%
   \ifVCShowState%
         \ifStateLineDbl\psset{style=VaucStateDblStyle}\else\psset{style=VaucStateStyle}\fi%
   \else\psset{style=VaucHiddenStateStyle}\fi}
\newcommand{\VaucStateRBLabel}[1]{%
    \textcolor{\StateLabelCol}{\scalebox{\StateLabelSca}{\scalebox{\StateLabelScale}{\rput[B]{0}(0,\VertShift){$ #1 $}}}}}%
\newcommand{\VaucStateLabel}[1]%
    {\ifVCShowState%
        \ifVCRigidLabel%
           \ifVCStateLabelBaseLine%
                 \textcolor{\StateLabelCol}{\scalebox{\StateLabelSca}{\scalebox{\StateLabelScale}{\rput[B]{*0}(0,\VertShift){$ #1 $}}}}%
           \else
                 \textcolor{\StateLabelCol}{\scalebox{\StateLabelSca}{\scalebox{\StateLabelScale}{\rput{*0}(0,0){$ #1 $}}}}%
           \fi
        \else
                 \textcolor{\StateLabelCol}{\scalebox{\StateLabelSca}{\scalebox{\StateLabelScale}{$ #1 $}}}%
        \fi
     \else%
                 \textcolor{white}{\scalebox{\StateLabelSca}{\scalebox{\StateLabelScale}{$ #1 $}}}%
     \fi}
\newcommand{\VCPutStateLabel}[2]%
    {\rput#1{\scalebox{\StateLabelSca}{$ #2 $}}}%
\newcommand{\ChgStateLineStyle}[1]{\renewcommand{\StateLineSty}{#1}}
\newcommand{\RstStateLineStyle}{\ChgStateLineStyle{\StateLineStyle}}
\newcommand{\SetStateLineStyle}[1]%
   {\renewcommand{\StateLineStyle}{#1}\RstStateLineStyle}%

\newcommand{\ChgStateLineWidth}[1]{\setlength{\StateLineWid}{#1\StateLineWidth}}%
\newcommand{\RstStateLineWidth}{\ChgStateLineWidth{1}}%
\newcommand{\SetStateLineWidth}[1]
   {\setlength{\StateLineWidth}{#1}\RstStateLineWidth}
\newcommand{\ChgStateLineColor}[1]{\renewcommand{\StateLineCol}{#1}}
\newcommand{\RstStateLineColor}{\ChgStateLineColor{\StateLineColor}}
\newcommand{\SetStateLineColor}[1]%
   {\renewcommand{\StateLineColor}{#1}\RstStateLineColor}
\newcommand{\ChgStateFillStatus}[1]{\renewcommand{\StateFillSta}{#1}}
\newcommand{\RstStateFillStatus}{\ChgStateFillStatus{\StateFillStatus}}
\newcommand{\SetStateFillStatus}[1]%
    {\renewcommand{\StateFillStatus}{#1}\RstStateFillStatus}
\newcommand{\ChgStateFillColor}[1]{\renewcommand{\StateFillCol}{#1}}
\newcommand{\RstStateFillColor}{\ChgStateFillColor{\StateFillColor}}
\newcommand{\SetStateFillColor}[1]%
    {\renewcommand{\StateFillColor}{#1}\RstStateFillColor}%
\newcommand{\ChgStateLabelColor}[1]{\renewcommand{\StateLabelCol}{#1}}
\newcommand{\RstStateLabelColor}{\ChgStateLabelColor{\StateLabelColor}}
\newcommand{\SetStateLabelColor}[1]%
    {\renewcommand{\StateLabelCol}{#1}\RstStateLabelColor}
\newcommand{\ChgStateLabelScale}[1]{\renewcommand{\StateLabelSca}{#1}}
\newcommand{\RstStateLabelScale}{\ChgStateLabelScale{1}}
\newcommand{\SetStateLabelScale}[1]%
   {\renewcommand{\StateLabelScale}{#1}\RstStateLabelScale}
\newcommand{\FixStateLineDouble}[2]{%
    \renewcommand{\StateLineDblCoef}{#1}%
    \renewcommand{\StateLineDblSep}{#2}}
\newcommand{\FixDimState}[5]{%
    \renewcommand{\DimStateLineStyle}{#1}%
    \renewcommand{\DimStateLineCoef}{#3}%
    \renewcommand{\DimStateLineColor}{#2}%
    \renewcommand{\DimStateLabelColor}{#4}%
    \renewcommand{\DimStateFillColor}{#5}}%
\newcommand{\RstState}{%
   \RstStateLineStyle\RstStateLineWidth%
   \RstStateLineColor%
   \RstStateFillStatus\RstStateFillColor%
   \RstStateLabelColor\RstStateLabelScale}%
\newlength{\StateDiam}
\newlength{\VaucAOS}\newlength{\VaucAOSdiag}
\newcommand{\StateSizeFlag}{}
\newcommand{\SetAOS}{%
   \setlength{\VaucAOS}{\ArrowOnStateCoef\StateDiam}%
   \setlength{\VaucAOSdiag}{\SQRTwo\VaucAOS}}
\newlength{\VariableStateIntDiam}
\newlength{\VariableStateWidth}
\newlength{\VariableStateITPos}
\newcommand{\SetStateIntDiam}{%
   \setlength{\VariableStateIntDiam}{\StateDiam}%
   \addtolength{\VariableStateIntDiam}{-2\StateLineWid}%
}%
\newcommand{\LoopSize}{}\newcommand{\LoopSi}{}
\newcommand{\LoopVarSize}{}\newcommand{\LoopVarSi}{}
\newcommand{\CLoopSize}{}\newcommand{\CLoopSi}{}
\newcommand{\ChgLoopSize}[1]{\renewcommand{\LoopSi}{#1}}
\newcommand{\RstLoopSize}{\ChgLoopSize{\LoopSize}}
\newcommand{\SetLoopSize}[1]%
   {\renewcommand{\LoopSize}{#1}\RstLoopSize}
\newcommand{\ChgCLoopSize}[1]{\renewcommand{\CLoopSi}{#1}}
\newcommand{\RstCLoopSize}{\ChgCLoopSize{\CLoopSize}}
\newcommand{\SetCLoopSize}[1]%
   {\renewcommand{\CLoopSize}{#1}\RstCLoopSize}
\newcommand{\ChgLoopVarSize}[1]{\renewcommand{\LoopVarSi}{#1}}
\newcommand{\RstLoopVarSize}{\ChgLoopVarSize{\LoopVarSize}}
\newcommand{\SetLoopVarSize}[1]%
   {\renewcommand{\LoopVarSize}{#1}\RstLoopVarSize}
\newcommand{\SetStateDiam}[4]{%
   \setlength{\StateDiam}{#1}%
   \renewcommand{\ArrowOnStateCoef}{#2}%
   \SetLoopSize{#3}%
   \SetLoopVarSize{#3}%
   \SetCLoopSize{#4}%
   \SetAOS\SetStateIntDiam}
\newcommand{\FixStateDiameter}[1]
   {\setlength{\StateDiam}{#1}\SetStateIntDiam \SetAOS}
\newcommand{\MediumState}%
   {\SetStateDiam{\MediumStateDiameter}{\ArrowOnMediumState}%
         {\LoopOnMediumState}{\CLoopOnMediumState}%
                  \renewcommand{\StateSizeFlag}{0}}
\newcommand{\SmallState}%
   {\SetStateDiam{\SmallStateDiameter}{\ArrowOnSmallState}%
         {\LoopOnSmallState}{\CLoopOnSmallState}%
                  \renewcommand{\StateSizeFlag}{1}}
\newcommand{\LargeState}%
   {\SetStateDiam{\LargeStateDiameter}{\ArrowOnLargeState}%
         {\LoopOnLargeState}{\CLoopOnLargeState}%
                  \renewcommand{\StateSizeFlag}{2}}
\newcommand{\RstStateSize}%
  {\ifthenelse{\equal{\StateSizeFlag}{0}}%
              {\MediumState}%
              {\ifthenelse{\equal{\StateSizeFlag}{1}}%
                              {\SmallState}{\LargeState}}}%
\newcommand{\VaucState}[3][{}]%
   {\rput#2{%
      \Cnode[radius=.5\StateDiam](0,0){#3}%
          \ifVCShowState%
      \nput[labelsep=-.5\StateDiam]{0}{#3}%
        {\makebox[0pt]{\VaucStateLabel{#1}}}%
      \fi
      \ifthenelse{\equal{\VCIFflag}{0}}{}{%
        \pnode(-\VaucAOS,0){#3w}\pnode(\VaucAOS,0){#3e}%
        \pnode(0,\VaucAOS){#3n}\pnode(0,-\VaucAOS){#3s}%
           \ifthenelse{\equal{\VCIFflag}{1}}{}{%
          \pnode(-\VaucAOSdiag,\VaucAOSdiag){#3nw}%
           \pnode(\VaucAOSdiag,\VaucAOSdiag){#3ne}%
           \pnode(-\VaucAOSdiag,-\VaucAOSdiag){#3sw}%
           \pnode(\VaucAOSdiag,-\VaucAOSdiag){#3se}%
                   }%
            }%
     }%
}
\newcommand{\State}[3][{}]{\StateStyle\VaucState[#1]{#2}{#3}}
\newcommand{\FinalState}[3][{}]%
   {\psset{style=VaucStateDblStyle}\VaucState[#1]{#2}{#3}}
\newcommand{\VSState}[2]%
    {\renewcommand{\ArrowOnStateCoef}{\ArrowOnVerySmallState}%
         \FixStateDiameter{\VerySmallStateDiameter}%
     \ChgStateLineWidth{\VSStateLineCoef}%
         \State{#1}{#2}%
         \RstStateLineWidth\RstStateSize}
\newlength{\ExtraSpace}
\newcommand{\StateVar}[3][]%
 {\StateStyle %
  \settowidth{\VariableStateWidth}{\scalebox{\StateLabelSca}{\scalebox{\StateLabelScale}{$#1$}}}%
  \addtolength{\VariableStateWidth}{\ExtraSpace}
  \ifthenelse{\lengthtest{\VariableStateWidth < \VariableStateIntDiam}}%
        {\setlength{\VariableStateWidth}{\VariableStateIntDiam}}{}%
  \setlength{\VariableStateITPos}{\ArrowOnStateCoef\StateDiam}%
  \addtolength{\VariableStateITPos}{0.5\VariableStateWidth}%
  \addtolength{\VariableStateITPos}{-0.5\StateDiam}%
  \rput#2{\pnode(\VariableStateITPos,0){#3e}%
          \pnode(-\VariableStateITPos,0){#3w}%
          \pnode(0,\ArrowOnStateCoef\StateDiam){#3n}%
          \pnode(0,-\ArrowOnStateCoef\StateDiam){#3s}}%
  \rput#2{\rnode{#3}{\psframebox{\protect\rule[-.5\VariableStateIntDiam]{0pt}{\VariableStateIntDiam}\protect\rule{\VariableStateWidth}{0pt}}}}
  \rput#2{\VaucStateRBLabel{#1}}%
}%

\newlength{\EdgeLineWid}
\newcommand{\EdgeLineSty}{\EdgeLineStyle}
\newcommand{\EdgeLineCol}{\EdgeLineColor}
\newcommand{\EdgeLabelSca}{1}
\newcommand{\EdgeLabelCol}{\EdgeLabelColor}
\newlength{\EdgeArrowSZDim}
\newcommand{\EdgeArrowSZNum}{\EdgeArrowLengthCoef}
\newcommand{\EdgeArrowSty}{\EdgeArrowStyle}
\newcommand{\EdgeArrowIns}{\EdgeArrowInset}
\newlength{\EdgeLineBord}
\newlength{\EdgeOff}
\newcommand{\VaucArcAng}{\VaucArcAngle}
\newcommand{\VaucLArcAng}{\VaucLArcAngle}
\newlength{\VaucArcOff}
\newcommand{\VaucArcCurv}{\VaucArcCurvature}
\newcommand{\VaucLArcCurv}{\VaucLArcCurvature}
\newcommand{\LoopAng}{\LoopAngle}
\newcommand{\CLoopAng}{\CLoopAngle}
\newcommand{\LoopVarAng}{\LoopVarAngle}
\newlength{\LoopOff}
\newlength{\LoopVarOff}
\newlength{\TransLabelSP}
\newcommand{\EdgeLabelPos}{\EdgeLabelPosit}
\newcommand{\ArcLabelPos}{\ArcLabelPosit}
\newcommand{\LArcLabelPos}{\LArcLabelPosit}
\newcommand{\LoopLabelPos}{\LoopLabelPosit}
\newcommand{\CLoopLabelPos}{\CLoopLabelPosit}
\newcommand{\InitStateLabelPos}{\InitStateLabelPosit}
\newcommand{\FinalStateLabelPos}{\FinalStateLabelPosit}
\newpsstyle{VaucEdgeStyle}%
    {arrows=\EdgeArrowSty,arrowsize=\EdgeArrowSZDim,arrowlength=\EdgeArrowSZNum,%
         arrowinset=\EdgeArrowIns,%
     linewidth=\EdgeLineWid,linecolor=\EdgeLineCol,linestyle=\EdgeLineSty,%
     doubleline=false,%
         bordercolor=\EdgeLineBorderColor,border=\EdgeLineBord,%
     fillstyle=none,offset=\EdgeOff,%
     labelsep=\TransLabelSP,nodesep=0pt}
\newpsstyle{VaucEdgeDblStyle}%
    {arrows=\EdgeArrowSty,arrowsize=\EdgeArrowSZDim,arrowlength=\EdgeArrowSZNum,%
         arrowinset=\EdgeArrowIns,%
     linewidth=\EdgeLineDblCoef\EdgeLineWid,linecolor=\EdgeLineCol,linestyle=\EdgeLineSty,%
     doubleline=true,doublesep=\EdgeLineDblSep\EdgeLineWid,%
         bordercolor=\EdgeLineBorderColor,border=\EdgeLineBord,%
     fillstyle=none,offset=\EdgeOff,%
     labelsep=\TransLabelSP,nodesep=0pt}
\newpsstyle{VaucArcR}{ncurv=\VaucArcCurv,arcangle=-\VaucArcAng,%
     labelsep=\TransLabelSP,offset=-\VaucArcOff}
\newpsstyle{VaucArcL}{ncurv=\VaucArcCurv,arcangle=\VaucArcAng,%
     labelsep=\TransLabelSP,offset=\VaucArcOff}
\newpsstyle{VaucLArcR}{ncurv=\VaucLArcCurv,arcangle=-\VaucLArcAng,%
     labelsep=\TransLabelSP,offset=-\VaucArcOff}
\newpsstyle{VaucLArcL}{ncurv=\VaucLArcCurv,arcangle=\VaucLArcAng,%
     labelsep=\TransLabelSP,offset=\VaucArcOff}
\newcommand{\EdgeStyle}{\ifEdgeLineDbl\psset{style=VaucEdgeDblStyle}%
        \else\psset{style=VaucEdgeStyle}\fi}
\newcommand{\ChgEdgeOffset}[1]{\setlength{\EdgeOff}{#1}}
\newcommand{\RstEdgeOffset}{\ChgEdgeOffset{\EdgeOffset}}
\newcommand{\SetEdgeOffset}[1]%
   {\setlength{\EdgeOffset}{#1}\RstEdgeOffset}

\newcommand{\ChgArcAngle}[1]{\renewcommand{\VaucArcAng}{#1}}
\newcommand{\RstArcAngle}{\ChgArcAngle{\VaucArcAngle}}
\newcommand{\SetArcAngle}[1]%
   {\renewcommand{\VaucArcAngle}{#1}\RstArcAngle}
\newcommand{\ChgLArcAngle}[1]{\renewcommand{\VaucLArcAng}{#1}}
\newcommand{\RstLArcAngle}{\ChgLArcAngle{\VaucLArcAngle}}
\newcommand{\SetLArcAngle}[1]%
   {\renewcommand{\VaucLArcAngle}{#1}\RstLArcAngle}
\newcommand{\ChgArcCurvature}[1]{\renewcommand{\VaucArcCurv}{#1}}
\newcommand{\RstArcCurvature}{\ChgArcCurvature{\VaucArcCurvature}}
\newcommand{\SetArcCurvature}[1]%
   {\renewcommand{\VaucArcCurvature}{#1}\RstArcCurvature}
\newcommand{\ChgLArcCurvature}[1]{\renewcommand{\VaucLArcCurv}{#1}}
\newcommand{\RstLArcCurvature}{\ChgLArcCurvature{\VaucLArcCurvature}}
\newcommand{\SetLArcCurvature}[1]%
   {\renewcommand{\VaucLArcCurvature}{#1}\RstLArcCurvature}

\newcommand{\RstArcOffset}{\setlength{\VaucArcOff}{\VaucArcOffset}}
\newcommand{\SetArcOffset}[1]%
   {\renewcommand{\VaucArcOffset}{#1}\RstArcOffset}

\newcommand{\RstLoopOffset}{\setlength{\LoopOff}{\LoopOffset}}
\newcommand{\SetLoopOffset}[1]%
   {\renewcommand{\LoopOffset}{#1}\RstLoopOffset}
\newcommand{\ChgLoopAngle}[1]{\renewcommand{\LoopAng}{#1}}
\newcommand{\RstLoopAngle}{\ChgLoopAngle{\LoopAngle}}
\newcommand{\SetLoopAngle}[1]%
   {\renewcommand{\LoopAngle}{#1}\RstLoopAngle}
\newcommand{\ChgCLoopAngle}[1]{\renewcommand{\CLoopAng}{#1}}
\newcommand{\RstCLoopAngle}{\ChgCLoopAngle{\CLoopAngle}}
\newcommand{\SetCLoopAngle}[1]%
   {\renewcommand{\CLoopAngle}{#1}\RstCLoopAngle}
\newcommand{\ChgEdgeLineColor}[1]{\renewcommand{\EdgeLineCol}{#1}}
\newcommand{\RstEdgeLineColor}{\ChgEdgeLineColor{\EdgeLineColor}}
\newcommand{\SetEdgeLineColor}[1]%
   {\renewcommand{\EdgeLineColor}{#1}\RstEdgeLineColor}
\newcommand{\ChgEdgeLineStyle}[1]{\renewcommand{\EdgeLineSty}{#1}}  
\newcommand{\RstEdgeLineStyle}{\ChgEdgeLineStyle{\EdgeLineStyle}}
\newcommand{\SetEdgeLineStyle}[1]%
   {\renewcommand{\EdgeLineStyle}{#1}\RstEdgeLineStyle}
\newcommand{\ChgEdgeLineWidth}[1]
   {\setlength{\EdgeLineWid}{#1\EdgeLineWidth}}
\newcommand{\RstEdgeLineWidth}{\ChgEdgeLineWidth{1}}
\newcommand{\SetEdgeLineWidth}[1]
   {\setlength{\EdgeLineWidth}{#1}\RstEdgeLineWidth}
\newcommand{\EdgeLineDouble}%
        {\EdgeLineDbltrue%
    \ChgEdgeArrowWidth{\EdgeDblArrowWidth}
    \ChgEdgeArrowLengthCoef{\EdgeDblArrowLengthCoef}}
\newcommand{\EdgeLineSimple}%
   {\EdgeLineDblfalse \RstEdgeArrowWidth \RstEdgeArrowLengthCoef}
\newcommand{\ChgEdgeLabelColor}[1]{\renewcommand{\EdgeLabelCol}{#1}}
\newcommand{\RstEdgeLabelColor}{\ChgEdgeLabelColor{\EdgeLabelColor}}
\newcommand{\SetEdgeLabelColor}[1]%
   {\renewcommand{\EdgeLabelColor}{#1}\RstEdgeLabelColor}
\newcommand{\ChgEdgeLabelScale}[1]{\renewcommand{\EdgeLabelSca}{#1}}
\newcommand{\RstEdgeLabelScale}{\ChgEdgeLabelScale{1}}
\newcommand{\SetEdgeLabelScale}[1]%
   {\renewcommand{\EdgeLabelScale}{#1}\RstEdgeLabelScale}
\newcommand{\FixDimEdge}[4]{%
    \renewcommand{\DimEdgeLineStyle}{#1}%
    \renewcommand{\DimEdgeLineCoef}{#2}%
    \renewcommand{\DimEdgeLineColor}{#3}%
    \renewcommand{\DimEdgeLabelColor}{#4}}%
\newcommand{\ChgEdgeArrowStyle}[1]{\renewcommand{\EdgeArrowSty}{#1}}
\newcommand{\RstEdgeArrowStyle}{\ChgEdgeArrowStyle{\EdgeArrowStyle}}
\newcommand{\SetEdgeArrowStyle}[1]%
   {\renewcommand{\EdgeArrowStyle}{#1}\RstEdgeArrowStyle}
\newcommand{\ChgEdgeArrowWidth}[1]%
   {\setlength{\EdgeArrowSZDim}{#1}} 
\newcommand{\RstEdgeArrowWidth}{\ChgEdgeArrowWidth{\EdgeArrowWidth}}
\newcommand{\SetEdgeArrowWidth}[1]%
   {\setlength{\EdgeArrowWidth}{#1} \RstEdgeArrowWidth}
\newcommand{\ChgEdgeArrowLengthCoef}[1]{\renewcommand{\EdgeArrowSZNum}{#1}}
\newcommand{\RstEdgeArrowLengthCoef}{\ChgEdgeArrowLengthCoef{\EdgeArrowLengthCoef}}
\newcommand{\SetEdgeArrowLengthCoef}[1]%
   {\renewcommand{\EdgeArrowLengthCoef}{#1}\RstEdgeArrowLengthCoef}
\newcommand{\ChgEdgeArrowInsetCoef}[1]{\renewcommand{\EdgeArrowIns}{#1}}
\newcommand{\RstEdgeArrowInsetCoef}{\ChgEdgeArrowInsetCoef{\EdgeArrowInset}}
\newcommand{\SetEdgeArrowInsetCoef}[1]%
   {\renewcommand{\EdgeArrowInset}{#1}\RstEdgeArrowInsetCoef}
\newcommand{\ReverseArrow}%
   {\ChgEdgeArrowStyle{\EdgeRevArrowStyle}%
    \renewcommand{\EdgeLabelPos}{\EdgeLabelRevPosit}%
    \renewcommand{\ArcLabelPos}{\ArcLabelRevPosit}%
    \renewcommand{\LArcLabelPos}{\LArcLabelRevPosit}%
    \renewcommand{\LoopLabelPos}{\LoopLabelRevPosit}%
    \renewcommand{\CLoopLabelPos}{\CLoopLabelRevPosit}%
    \renewcommand{\InitStateLabelPos}{\InitStateLabelRevPosit}%
    \renewcommand{\FinalStateLabelPos}{\FinalStateLabelRevPosit}}
\newcommand{\StraightArrow}%
   {\ChgEdgeArrowStyle{\EdgeArrowStyle}%
    \renewcommand{\EdgeLabelPos}{\EdgeLabelPosit}%
    \renewcommand{\ArcLabelPos}{\ArcLabelPosit}%
    \renewcommand{\LArcLabelPos}{\LArcLabelPosit}%
    \renewcommand{\LoopLabelPos}{\LoopLabelPosit}%
    \renewcommand{\CLoopLabelPos}{\CLoopLabelPosit}%
    \renewcommand{\InitStateLabelPos}{\InitStateLabelPosit}%
    \renewcommand{\FinalStateLabelPos}{\FinalStateLabelPosit}}
\newcommand{\FixEdgeLineDouble}[2]{%
    \renewcommand{\EdgeLineDblCoef}{#1}%
    \renewcommand{\EdgeLineDblSep}{#2}}
\newcommand{\FixEdgeBorder}[2]{%
    \renewcommand{\EdgeLineBorderCoef}{#1}%
    \renewcommand{\EdgeLineBorderColor}{#2}}
\newcommand{\EdgeBorder}%
  {\setlength{\EdgeLineBord}{\EdgeLineBorderCoef\EdgeLineWid}}

\newcommand{\VaucEdgeLabel}[1]{%
        \textcolor{\EdgeLabelCol}{\scalebox{\EdgeLabelSca}{\scalebox{\EdgeLabelScale}{$ #1 $}}}}
\newcommand{\RstEdge}{%
   \RstEdgeOffset\RstArcAngle\RstLArcAngle%
   \RstArcCurvature\RstLArcCurvature%
   \RstArcOffset\RstLoopOffset\RstLoopSize%
   \RstEdgeLineColor\RstEdgeLineStyle\RstEdgeLineWidth\EdgeLineSimple%
   \StraightArrow
   \RstEdgeLabelScale\RstEdgeLabelColor}

\newcommand{\Initial}[2][\InitialDir]{\EdgeStyle\ncline{#2#1}{#2}}
\newcommand{\Final}[2][\FinalDir]{\EdgeStyle\ncline{#2}{#2#1}}
\newcommand{\InitialL}[4][{\InitStateLabelPos}]%
   {\EdgeStyle\ncline{#3#2}{#3}\naput[npos=#1]{\VaucEdgeLabel{#4}}}
\newcommand{\InitialR}[4][{\InitStateLabelPos}]%
   {\EdgeStyle\ncline{#3#2}{#3}\nbput[npos=#1]{\VaucEdgeLabel{#4}}}
\newcommand{\FinalL}[4][{\FinalStateLabelPos}]%
   {\EdgeStyle\ncline{#3}{#3#2}\naput[npos=#1]{\VaucEdgeLabel{#4}}}
\newcommand{\FinalR}[4][{\FinalStateLabelPos}]%
   {\EdgeStyle\ncline{#3}{#3#2}\nbput[npos=#1]{\VaucEdgeLabel{#4}}}
\newcommand{\EdgeL}[4][{\EdgeLabelPos}]%
   {\EdgeStyle \ncline{#2}{#3} \naput[npos=#1]{\VaucEdgeLabel{#4}}}
\newcommand{\EdgeR}[4][{\EdgeLabelPos}]%
   {\EdgeStyle \ncline{#2}{#3} \nbput[npos=#1]{\VaucEdgeLabel{#4}}}
\newcommand{\ArcL}[4][{\ArcLabelPos}]%
   {\EdgeStyle \psset{style=VaucArcL}%
    \ncarc{#2}{#3} \naput[npos=#1]{\VaucEdgeLabel{#4}}}
\newcommand{\ArcR}[4][{\ArcLabelPos}]%
   {\EdgeStyle \psset{style=VaucArcR}%
    \ncarc{#2}{#3} \nbput[npos=#1]{\VaucEdgeLabel{#4}}}
\newcommand{\LArcL}[4][{\LArcLabelPos}]%
   {\EdgeStyle \psset{style=VaucLArcL}%
    \ncarc{#2}{#3} \naput[npos=#1]{\VaucEdgeLabel{#4}}}
\newcommand{\LArcR}[4][{\LArcLabelPos}]%
   {\EdgeStyle \psset{style=VaucLArcR}%
    \ncarc{#2}{#3} \nbput[npos=#1]{\VaucEdgeLabel{#4}}}
\newcounter{anglea}\newcounter{angleb}
\newcommand{\LoopXR}[7]%
   {{\setcounter{anglea}{#2-#4}}%
    {\setcounter{angleb}{#2+#4}}%
    {\EdgeStyle \psset{angleA=\theanglea,angleB=\theangleb,offset=#5,ncurv=#6}%
    \nccurve{#3}{#3} \nbput[npos=#1]{\VaucEdgeLabel{#7}}}}
\newcommand{\LoopXL}[7]%
   {{\setcounter{anglea}{#2+#4}}%
    {\setcounter{angleb}{#2-#4}}%
    {\EdgeStyle \psset{angleA=\theanglea,angleB=\theangleb,offset=-#5,ncurv=#6}%
    \nccurve{#3}{#3} \naput[npos=#1]{\VaucEdgeLabel{#7}}}}
\newcommand{\LoopR}[4][{\LoopLabelPos}]%
   {\LoopXR{#1}{#2}{#3}{\LoopAng}{\LoopOff}{\LoopSi}{#4}}
\newcommand{\LoopL}[4][{\LoopLabelPos}]%
   {\LoopXL{#1}{#2}{#3}{\LoopAng}{\LoopOff}{\LoopSi}{#4}}
\newcommand{\CLoopR}[4][{\CLoopLabelPos}]%
   {\LoopXR{#1}{#2}{#3}{\CLoopAng}{\LoopOff}{\LoopSi}{#4}}
\newcommand{\CLoopL}[4][{\CLoopLabelPos}]%
   {\LoopXL{#1}{#2}{#3}{\CLoopAng}{\LoopOff}{\LoopSi}{#4}}
\newcommand{\LoopVarR}[4][{\LoopLabelPos}]%
   {\LoopXR{#1}{#2}{#3}{\LoopVarAng}{\LoopVarOff}{\LoopVarSi}{#4}}
\newcommand{\LoopVarL}[4][{\LoopLabelPos}]%
   {\LoopXL{#1}{#2}{#3}{\LoopVarAng}{\LoopVarOff}{\LoopVarSi}{#4}}
\newcommand{\LoopW}[3][{\LoopLabelPos}]{\LoopR[#1]{180}{#2}{#3}}

\newcommand{\LoopN}[3][{\LoopLabelPos}]{\LoopL[#1]{90}{#2}{#3}}
\newcommand{\LoopS}[3][{\LoopLabelPos}]{\LoopR[#1]{-90}{#2}{#3}}

\newcommand{\LoopSE}[3][{\LoopLabelPos}]{\LoopR[#1]{-45}{#2}{#3}}

\newcommand{\LoopVarS}[3][{\CLoopLabelPos}]{\LoopVarR[#1]{-90}{#2}{#3}}
\newcommand{\VArcR}[5][{\ArcLabelPos}]%
   {\EdgeStyle \psset{style=VaucLArcR}%
    \ncarc[#2]{#3}{#4} \nbput[npos=#1]{\VaucEdgeLabel{#5}}}
\SetStateLabelColor{black}              
\SetStateLabelScale{1.7}                
\SetStateLineStyle{solid}               
\SetStateLineWidth{1.8pt}               
\SetStateLineColor{black}               
\SetStateFillStatus{solid}              
\SetStateFillColor{white}               
\FixDimState{solid}{gray}{1}{gray}{white}  
\FixStateLineDouble{0.6}{0.4}               
\SetEdgeLabelColor{black}               
\SetEdgeLabelScale{1.7}                 
\SetEdgeLineStyle{solid}                
\SetEdgeLineWidth{1pt}                  
\SetEdgeLineColor{black}                
\SetArcAngle{15}                        
\SetLArcAngle{30}                       
\SetArcCurvature{0.8}                   
\SetEdgeOffset{0pt}                     
\SetArcOffset{1pt}                      
\SetLoopOffset{0pt}                     
\FixDimEdge{solid}{1.2}{gray}{gray}     
\FixEdgeBorder{2}{white}                
\FixEdgeLineDouble{0.4}{0.8}            
\renewcommand{\VSStateLineCoef}{.6}             
\SetEdgeArrowWidth{5pt}         
\SetEdgeArrowLengthCoef{1.4}            
\renewcommand{\EdgeDblArrowLengthCoef}{1.7}     
\SetEdgeArrowInsetCoef{0.1}                     
\SetEdgeArrowStyle{->}                          
\renewcommand{\EdgeRevArrowStyle}{<-}           
\renewcommand{\StateDimen}{outer}               
\renewcommand{\StateDblDimen}{middle}           
\SetVCDirectory{}        
\MediumPicture
\FullState                              
\MediumState
\renewcommand{\cal}{\mathcal}

\newcommand{\supp}[1]{\text{Supp }#1}

\renewcommand\leq{\leqslant} 
\renewcommand\geq{\geqslant} 
\def\Twin{\text{Twin}}
\newcommand{\moins}{\!\setminus \!}

\newcommand{\avw}[3]{\textit{av}\left(#1, #2, #3\right)}

\def\cA{{\mathcal A}}

\def\cR{{\mathcal R}}
\def\cP{{\mathcal P}}

\newcommand{\D}{\mathcal{D}}
\newcommand{\Bc}{\mathcal{B}}

\newcommand{\Sc}{\mathcal{S}}
\newcommand{\Uc}{\mathcal{U}}
\newcommand{\Ac}{\mathcal{A}}

\newcommand{\diagramm}[2]{\begin{array}[t]{c}
#1\\[-.2cm]
{\tiny \text{(\S{#2})}}
\end{array}
}
\newcommand{\tourne}[2]{\text{\begin{rotate}{#2}
${#1}$
\end{rotate}}
}
\newcommand{\N} {\ensuremath{\mathbb{N}}}
\newcommand{\Z} {\ensuremath{\mathbb{Z}}}

\newcommand{\R} {\ensuremath{\mathbb{R}}}
\newcommand{\K} {\ensuremath{\mathbb{K}}}
\newcommand{\B} {\ensuremath{\mathbb{B}}}
\newcommand{\Nmin} {\N_{\min}}
\newcommand{\Nmax} {\N_{\max}}

\newcommand{\Zmax} {\Z_{\max}}

\newcommand{\Rmin} {\R_{\min}}
\newcommand{\Rmax} {\R_{\max}}
\def\plus{\oplus}
\def\bigplus{\bigoplus}
\def\fois{\otimes}
\def\bigfois{\bigotimes}
\def\vide{\varepsilon}
\def\emptyword{\varepsilon}
\newcommand{\1}{\mathbb{1}}
\newcommand{\0}{\mathbb{0}}
\def\zero{\0}
\def\un{\1}

\newcommand{\flech}[1]{\overset{#1}{\longrightarrow}}
\newcommand{\fleche}[2]{\mathchoice
         {\xrightarrow{#1\mid #2}}
         {\xrightarrow{\smash{\lower1pt\hbox{$\scriptstyle #1$}}}}
         {\text{Erreur}}
         {\text{Erreur}}}
\newcommand{\chemin}[2]{\Bigl[#1\Bigr]_{#2}}

\newcommand{\wght}[1]{\textit{weight}\left(#1\right)}

\newcommand{\init}[2][]{\overset{#1}{\rightarrow} {#2}}
\newcommand{\final}[2][]{{#2} \overset{#1}{\rightarrow}}
\newcommand{\series}[2]{#1\langle\! \langle #2 \rangle\! \rangle}
\newcommand{\coef}[2]{\langle #1, #2\rangle}

\newcommand{\set}[2]{\{#1\mid\;#2\}}
\renewcommand\emptyset{\varnothing}
\def\Rat{\text{Rat}}
\def\FAmb{\text{FAmb}}
\def\FUSeq{\text{FSeq}}
\def\NAmb{\text{NAmb}}
\def\Seq{\text{Seq}}
\def\Lip{\text{Lip}}
\def\Ser{\text{Series}}
\def\ab{\Sigma}

\newcommand{\Past}[2][\A]{\mathsf{Past}_{#1}(#2)}
\newcommand{\Fut}[2][\A]{\mathsf{Fut}_{#1}(#2)}
\def\A{\mathcal{A}}
\def\Ua{\mathcal{U}} 
\def\P{\mathcal{P}} 
\newcommand{\ind}[1]{^{\llcorner\raisebox{.4ex}{\tiny\!\!$#1$}}}
\newcommand{\p}[1]{\mathbf{#1}}
\newcommand{\vict}[1]{\textit{Vict}\left(#1\right)}

\def\ie{{\em i.e.\ }}
\def\oo{\infty}
\def\x{\times}
\newcommand{\vmin}[1]{\check{#1}}
\newcommand{\vnorm}[1]{\underline{#1}}
\newtheorem{prop}{Proposition}
\newtheorem{thm}{Theorem}
\newtheorem{defn}{Definition}
\newtheorem{lem}{Lemma}
\newtheorem{cor}{Corollary}
\newenvironment{pf}{{\bf proof.}}{}
\newcommand{\qed}{{~}\hfill$\square$}
\begin{document}
\title{Deciding Unambiguity and Sequentiality starting from a Finitely
  Ambiguous Max-Plus Automaton}
\author{Ines Klimann, Sylvain Lombardy,
Jean Mairesse,\\ and Christophe Prieur\thanks{%
{\sc LIAFA}, CNRS ({\sc umr 7089}) - Universit\'e
Paris 7, 2, place Jussieu - 75251 Paris Cedex 5 - France.
email: {\small{\tt\{klimann,lombardy,mairesse,prieur\}@liafa.jussieu.fr}}}}
\date{July 4, 2004}

\maketitle

\begin{abstract}
  Finite automata with weights in the max-plus semiring are
  considered. The main result is: it is decidable in an effective way
  whether a series that is recognized by a finitely ambiguous max-plus
  automaton is unambiguous, or is sequential. A collection of examples
  is given to illustrate the hierarchy of max-plus series with respect
  to ambiguity.
\end{abstract}

\section{Introduction}

A {\em max-plus automaton} is a finite automaton with multiplicities in the
max-plus semiring $\Rmax=(\R\cup\{-\infty\},\max,+)$. 
Roughly speaking, it is an automaton with two tapes: an input tape
labelled by a finite alphabet $\ab$, and an output tape weighted in
$\Rmax$. The weight of a word in $\ab^*$ is the maximum over all
successful paths of the sum of the weights along the path. 

Max-plus automata, and their min-plus counterparts, are studied under
various names in the literature: distance automata, finance automata,
cost automata. They have also appeared in various contexts: to study
logical problems in formal language theory (star height, finite power
property)~\cite{hash88,simo88}, to model the dynamic of some Discrete
Event Systems (DES)~\cite{gaub93,GaMa98b}, or in the context of automatic
speech recognition~\cite{mohr}.

Two automata are {\em equivalent} if they recognize the same series,
\ie if they have the same input/output behavior. The problem of
equivalence of two max-plus automata is undecidable~\cite{krob}. The
same problem for finitely ambiguous max-plus automata is decidable
\cite{HIJi,w94}.

The {\em sequentiality} problem is defined as follows: given a max-plus
automaton, is there an equivalent max-plus automaton which is {\em sequential}
(\ie deterministic in input). Let us give some motivations on
why the sequentiality problem is important. In the case of a
sequential automaton, the time complexity of computing the output is
roughly linear in the length of the input. This time efficiency is
central in speech processing, see~\cite{mohr}. Consider now a DES modelled
by a max-plus automaton. If the automaton is unambiguous, or
{\it a fortiori} sequential, then one can compute the optimal, as well as
the average behavior, of the DES, see~\cite{gaub93,GaMa95}.

\noindent Sequentiality is decidable for unambiguous max-plus automata
\cite{mohr}. 
In the present paper, we prove that sequentiality is decidable for
finitely ambiguous max-plus automata. 
To the best of our knowledge, it is not known if the finite ambiguity
of a max-plus series (defined {\it via} an infinitely ambiguous
automaton) is a decidable problem. In particular, the status of the
sequentiality problem is
still open for a general max-plus automaton (even if the multiplicities
are restricted to be in $\Zmax$, $\Nmax$ or $\Z^{-}_{\max}$).  
To be complete, it is necessary to mention that in~\cite[\S 3.5]{mohr}, 
it is claimed that any
max-plus automaton admits an effectively computable equivalent
unambiguous one. If that was true, it
would imply the decidability of the sequentiality for general max-plus
automata. However, the statement is erroneous and counter-examples are
provided in \S\ref{sec:Hier} of the present paper\footnote{The version
  of~\cite{mohr} available on the author's website has been correctly
  modified.}.

The sequentiality problem can be asked for automata over any semiring
$\K$. For transducers, \ie when $\K$ is the set of rational subsets of
a free monoid (with union and concatenation as the two laws), the
problem is completely solved in the functional case (when, for every
input, the output is a language of cardinality at most one)
\cite{bers79,chof77,chof}. For a general transducer, the problem is 
wide open. Observe that the  semiring $\{a^{\geq n},n\in
\N\}=\{a^na^*,n\in \N\}$ is isomorphic to $\N_{\min}$: $a^na^* +
a^ma^*=a^{\min(n,m)}a^*$ and $a^na^* \cdot 
a^ma^*=a^{n+m}a^*$. Similarly,  the semiring $\{a^{\leq n},n\in \N\}$
 is isomorphic to $\N_{\max}$ (where $a^{\leq n}=\{\vide, a,\ldots
 ,a^n\}$). Hence automata over $\Nmax$ or $\Nmin$ translate into
 transducers, but {\em not} functional ones. Also
 the translation does not work for automata over $\Rmax$. Hence, the
 vast literature on transducers is of limited use in our context.

In the present paper, we work with $\Rmax$. Decidability and
complexity should be interpreted under the  assumption that two real
numbers can be added or compared in constant time. 

\section{Preliminaries}
\subsection{Max-plus semiring and series}
The free monoid over a finite set (alphabet) $\ab$ is denoted by
$\ab^*$ and the empty word is denoted by~$\emptyword$. 
The structure
$\Rmax=(\R\cup\{-\infty\},\max,+)$ is a semiring, which
is called the {\em max-plus semiring}. 
It is convenient to use the notations $\oplus=\max$ and $\otimes=+$. 
The neutral elements of $\oplus$ and $\otimes$ are denoted
respectively by $\0=-\infty$ and $\1=0$.  
The subsemirings $\Nmax$, $\Zmax$, \ldots, are defined in the natural
way. The {\em min-plus semiring} $\R_{\min}$ is obtained by replacing
$\max$ by $\min$ and $-\infty$ by $+\infty$
in the definition of $\Rmax$. The results of this paper
can be easily adapted to the min-plus setting.
Observe that the subsemiring $\B=(\{\0,\1\},\oplus,\otimes)$ is
isomorphic to the Boolean semiring. 
For matrices $A,B$ of appropriate sizes with
entries in $\Rmax$, we set
$(A\oplus B)_{ij} =A_{ij} \oplus B_{ij}$,
$(A\otimes B)_{ij} =\bigoplus_{k} A_{ik} \otimes B_{kj}$,
and for $a\in \Rmax$, $(a\otimes A)_{ij} =a\otimes
A_{ij}$.  We usually omit the $\otimes$ sign,
writing for instance $AB$ instead
of $A\otimes B$. 

Consider the set $\series{\Rmax}{\ab^*}$ of {\em (formal power)
  series (over $\ab^*$ with coefficients in $\Rmax$)}, that is the set
  of maps from $\ab^*$ to $\Rmax$. We denote by $\coef{S}{u}$ the
  coefficient of the word $u$ in the series $S$.
  The {\em support} of a series $S$ is the set
  $\supp{S}=\set{u\in \ab^*}{\coef{S}{u}\neq \0}$.  
It is convenient to use the notation 
$S=\bigoplus_{u\in \ab^*}
  \coef{S}{u} u= \bigoplus_{u\in \supp{(S)}} \coef{S}{u} u$.
Equipped with the addition ($\oplus$) and
the Cauchy product ($\otimes$), the set $\series{\Rmax}{\ab^*}$ forms a semiring. 
The image of $\lambda\in \Rmax$ by the canonical injection into
  $\series{\Rmax}{\ab^*}$ is still denoted by $\lambda$. In
  particular, the neutral elements of $\series{\Rmax}{\ab^*}$ are $\0$
  and $\1$. The characteristic series of a language $L$ is the series
  $\1_L$ such that $\coef{\1_L}{w}=\1$ if $w\in L$, and
  $\coef{\1_L}{w}=\0$ otherwise.

\subsection{Max-plus automaton}
Let $Q$ and $\ab$ be two finite sets.
A {\em max-plus automaton} of set of states (dimension) $Q$ over the alphabet $\ab$, 
is a triple $\cA= (\alpha,\mu,\beta)$, where $\alpha\in \Rmax^{1\times Q}$,
$\beta\in \Rmax^{Q\times 1}$,
and where $\mu: \ab^* \rightarrow \Rmax^{Q\times Q}$ is a
morphism of monoids.
The morphism $\mu$ is uniquely determined by the family
of matrices $\{\mu(a),a \in \ab \},$ and for $w=a_1\cdots a_n$, we have
$\mu(w)=\mu(a_1)\otimes \cdots \otimes \mu(a_n)$. 
The series {\em recognized\/} (or {\em realized}) {\em by} $\A$ is
  by definition $S(\A)=\bigoplus_{u\in \ab^*} (\alpha\mu(u)\beta) u$.  
This is just a specialization to the max-plus semiring of the
classical notion of an automaton with multiplicities over a semiring
\cite{BeRe,eile,KS}.  By the Kleene-Sch\"utzenberger
Theorem~\cite{Sch61}, the set of series 
  recognized by  a max-plus automaton is equal to the set of rational
  series over $\Rmax$. We denote it by $\Rat$.

A state $i\in Q$ is {\em initial}, resp. {\em final}, if $\alpha_i \neq \0$,
resp. $\beta_i \neq \0$.  
As usual a max-plus automaton is represented graphically by a labelled
weighted digraph with ingoing and outgoing arcs for initial and final
states, see {\it e.g.} Figure \ref{fig:NA/nFUS1} (the input or output
weights equal to $\un$ are omitted).  
The terminology of graph theory is used accordingly ({\it e.g.} (simple) path or
circuit of an automaton, union of automata, \ldots).
A path which is both starting with an ingoing arc and 
ending with an outgoing arc is called a {\em successful path}.  
The {\em label of a path} is the concatenation of the labels of the
successive arcs (so called {\em transitions}), the {\em weight of a 
  path} is the product ($\otimes$) of the weights of the successive
arcs (including the ingoing and the outgoing arc, need it be).
We denote by $\wght{\pi}$ the weight of the path $\pi$. 
We use the following notations for paths in an automaton
  $\A=(\alpha,\mu,\beta)$: \[
  p\rightarrow q, \ \ \rightarrow p\rightarrow q, \ \ p\rightarrow q
  \rightarrow, \ \ p\fleche{u}{x}q, \ \ \chemin{p\fleche{u}{x}q}{\cA}\:,
  \text{ if } \mu(u)_{pq}=x \text{ in } \A \:. \]

The first example is a path (of any length) from $p$ to $q$, the
second also includes an ingoing arc, the third an outgoing arc, in the
fourth the weight and the label are added and in the fifth the
underlying automaton is recalled.

An automaton is {\em trim} if any state belongs to at least one successful path. 

Let $I$ be a finite set.
The {\em tensor product
  automaton} of $(\A_i=(\alpha\ind{i},
\mu\ind{i}, \beta\ind{i}))_{i\in I}$, denoted by $\odot_{i\in I} \A_i$, is defined  
as follows. It is the max-plus
automaton $(A, M, B)$ of 
dimension $Q=\prod_i Q_i$, where $Q_i$ is the dimension of $\cA_i$,
  and such that
\begin{equation*}
\forall p,q \in Q, \  \ A_{p} = \bigfois_{i\in I} \alpha\ind{i}_{p_i}, \ \ \forall a\in \ab, \
M(a)_{p,q}= \bigfois_{i\in I} \mu\ind{i}(a)_{p_i,q_i}, \ \ B_{p}=
\bigfois_{i\in I} \beta\ind{i}_{p_i}\:.
\end{equation*}

\subsection{Heap model}
A {\em heap} or {\em Tetris model}~\cite{vien}, consists of a finite
set of slots $\cR$, and a finite set of rectangular pieces $\ab$. Each
piece $a\in \ab$ is of height 1 and occupies a determined subset
$\cR(a)$ of the slots. To a word $u=u_1\cdots u_k\in \ab^*$ is
associated the {\em heap} obtained by piling up in order the pieces
$u_1,\dots, u_k$, starting with a horizontal ground and according to
the Tetris game mechanism (pieces are subject to gravity and fall down
vertically until they meet either a previously piled up piece or the
ground).  Consider the morphism generated by the matrices $M(a)\in
\R_{\max}^{\cR\times\cR}, a\in \ab,$ defined by
\\\hspace*{4cm}$\displaystyle{
M(a)_{ij} = \left\{\begin{array}{ll}
1 & \text{ if } i,j \in \cR(a),\\ 0 & \text{ if }
i=j \not\in \cR(a),\\-\infty &\text{ otherwise}\:.
\end{array}\right.
}$
\\Let $x(u)_i$ be the
height of the heap $u$ on slot $i\in \cR$. 
We have (\cite{BrVi,GaMa95,GaMa98b}): $x(u)_i=\1 M(u)\delta_i$, where
$\1=(\1,\dots, \1)\in \Rmax^{1\times \cR}$ and $\delta_i \in
\Rmax^{\cR\times 1}$ is defined by $(\delta_i)_j = \1$ if $j=i$ and
$\0$ otherwise. In other words, the application $x(\cdot)_i: \ab^*
\rightarrow \Rmax$ is recognized by the max-plus automaton $(\1,M,\delta_i)$. 
We call $(\1,M,\delta)$, $\delta=\bigoplus_{i\in I}\delta_i, I\subseteq
\cR$, a {\em heap automaton} (associated with the heap model). 
Among max-plus automata, heap automata are particularly convenient and
playful, due to the underlying geometric interpretation. Here, they
are used as a source of examples and counter-examples, {\it e.g.}
Figures~\ref{fig:FUS/nNA}, \ref{fig:NA/nFUS0} and~\ref{fig:nNA/nFUS/FA2}. 

\begin{figure}[htbp]
  \centering
\begin{picture}(0,0)%
\includegraphics{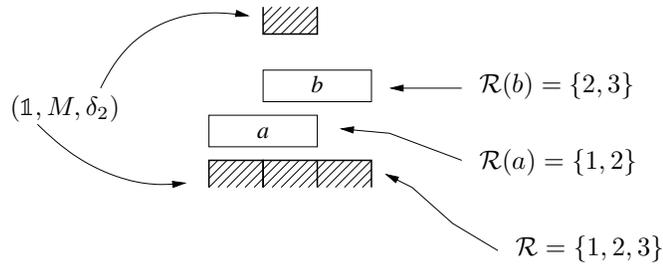}%
\end{picture}%
\setlength{\unitlength}{2486sp}%
\begingroup\makeatletter\ifx\SetFigFont\undefined%
\gdef\SetFigFont#1#2#3#4#5{%
  \reset@font\fontsize{#1}{#2pt}%
  \fontfamily{#3}\fontseries{#4}\fontshape{#5}%
  \selectfont}%
\fi\endgroup%
\begin{picture}(5040,2564)(2341,-5123)
\put(7021,-4201){\makebox(0,0)[lb]{\smash{{\SetFigFont{10}{12.0}{\familydefault}{\mddefault}{\updefault}{\color[rgb]{0,0,0}${\cal R}(a)=\{1,2\}$}%
}}}}
\put(7021,-3436){\makebox(0,0)[lb]{\smash{{\SetFigFont{10}{12.0}{\familydefault}{\mddefault}{\updefault}{\color[rgb]{0,0,0}${\cal R}(b)=\{2,3\}$}%
}}}}
\put(7381,-5056){\makebox(0,0)[lb]{\smash{{\SetFigFont{10}{12.0}{\familydefault}{\mddefault}{\updefault}{\color[rgb]{0,0,0}${\cal R}=\{1,2,3\}$}%
}}}}
\put(2341,-3661){\makebox(0,0)[lb]{\smash{{\SetFigFont{10}{12.0}{\rmdefault}{\mddefault}{\updefault}{\color[rgb]{0,0,0}$(\1,M,\delta_2)$}%
}}}}
\end{picture}%
\caption{A heap automaton}
  \label{fig-tas}
\end{figure}

We represent a heap automaton graphically as in
Figure~\ref{fig-tas}.

\subsection{Ambiguity and Sequentiality}\label{sse-as}

Consider a max-plus automaton $\A=(\alpha,\mu,\beta)$ of dimension $Q$
over $\ab$. The automaton is {\em sequential} if  there is a unique
initial state  and if for all $i\in Q$, and for all $a\in \ab$, there is at
most one $j\in Q$ such that $\mu(a)_{ij}\neq \0$. 
In the case of a Boolean automaton, we also say {\em deterministic} for
sequential. The automaton $\A$ is {\em unambiguous} if for any word
$u\in \ab^*$, there is at most one successful path of label $u$. 
The automaton is {\em finitely ambiguous} 
if there exists some $k\in \N$
such that for any word
$u\in \ab^*$, there are at most $k$ successful paths of label $u$. 
The minimal such $k$ is called the {\em degree of ambiguity} of the automaton. 
Clearly, `sequential' implies `unambiguous' which implies `finitely
ambiguous'. 
The automaton is {\em infinitely ambiguous} if it is not finitely ambiguous. 

Consider a series $S \in \Rat$. The series is {\em sequential}
  (resp. {\em unambiguous}, {\em finitely ambiguous}) if there exists
  a sequential (resp. unambiguous, finitely ambiguous) max-plus
  automaton recognizing it. The series is {\em infinitely ambiguous}
  if there exists no finitely ambiguous max-plus automaton
  recognizing it.
The {\em
  degree of ambiguity} of a finitely ambiguous series is the minimal
  degree of ambiguity of an automaton recognizing it. 
The sets of sequential, unambiguous, and finitely ambiguous series
  are denoted respectively by $\Seq$, $\NAmb$, and $\FAmb$. 
Define $\FUSeq = \set{ S }{\exists k, \exists S_1,\dots, S_k \in \Seq,
  \ S=S_1\oplus \cdots \oplus S_k}$.

Consider a total order on $\ab^*$. Given a series $S\neq \0$,
define the {\em normalized series} $\varphi(S)$ by 
$\varphi(S)= \bigoplus_{u\in \ab^*} (\coef{S}{u} -
  \coef{S}{u_0}) u$,
where $u_0$ is the smallest word of $\supp{S}$. 
The {\em (left)
quotient} of a series $S$ by a word $w$ is the series $w^{-1}S$ defined
by
$w^{-1}S= \bigoplus_{u\in \ab^*} \coef{S}{wu} u$.

A series $S$ is rational if and only if the semi-module of series $
\langle w^{-1}S, w\in \ab^* \rangle$ is finitely generated, \ie if
there exists $S_1,\dots ,S_k,$ such that: $$\forall w\in \ab^*,\, \exists
\lambda_1,\dots , \lambda_k \in \Rmax, \ w^{-1}S  = \bigoplus_i
\lambda_i S_i.$$
A series $S$ is sequential if and only if the set of series $\{
\varphi(w^{-1}S), w\in \ab^* \}$ is finite. 

\begin{prop}
A trim automaton
$\cA$ of dimension $Q$ is infinitely ambiguous if and only if there
exist $p,q\in Q, p\neq q,$ and $v\in \ab^*$, such that 
\mbox{$p\flech{v} p$}, \mbox{$p\flech{v}q$}, \mbox{$q\flech{v}q$}.
This can be checked in polynomial time.
\end{prop}

For a proof, see~\cite{WeSe} and the references therein. 
Observe that the (in)finite ambiguity is independent of the underlying semiring. 
Next result is due to Mohri~\cite{mohr} and is an adaptation  of a classical
result of Choffrut on functional transducers, see 
\cite{bers79,chof77,chof} (for the decidability) and
\cite{BCPS,WeKl} (for the polynomial complexity). 

\begin{thm}
\label{th-chof}
Let $\A$ be an unambiguous max-plus automaton. There exists a
polynomial time algorithm to decide whether $S(\A)$ is a sequential
series. 
\end{thm}

If $\A$ is unambiguous and $S(\A)$ is sequential, a sequential automaton recognizing
the series can be effectively constructed from $\A$ using an
adaptation of the subset construction of Boolean automata~\cite{MA02,BGW00,mohr}.



\bigskip
It is useful to detail Theorem \ref{th-chof}. 
We need to introduce several definitions. 
Given two words $u,v\in \ab^*$, let $u\wedge v$ be the longest common
prefix of $u$ and $v$, and define $d(u,v)=|u|+|v| -2|u\wedge v|$. It
is easy to check that $d(.,.)$ is a distance on $\ab^*$. A series $S$
is {\em $M$-Lipschitz} ($M\in \R_+$) if: 
\[
\forall u,v \in \supp{S}, \ 
|\coef{S}{u}-\coef{S}{v}| \leq M d(u,v)\:;
\]
and $S$ is {\em Lipschitz} if it
is $M$-Lipschitz for some $M$. The set of Lipschitz series is denoted
by $\Lip$. 
Consider a trim max-plus automaton $\A$ of dimension $Q$. Two states
$p,q\in Q$ are {\em twins} if:
\[
\Bigl[ \flech{x_0} i \fleche{u_1}{x_1}p \fleche{u_2}{x_2} p , \ \ \ \flech{y_0} j
\fleche{u_1}{y_1}q \fleche{u_2}{y_2} q \Bigr] \implies [ x_2=y_2]\:.
\]
If all the
states are twins, the automaton $\A$ is said to satisfy the
{\em twin property}. We denote the set of all such automata by $\Twin$. 
The following implications hold: 
\begin{equation}\label{eq-chof}
\Bigl[ \A \in \Twin \Bigr] \ \implies \ \Bigl[ S(\A) \in \Seq \Bigr] \
\implies \  \Bigl[ S(\A) \in \Lip \Bigr]\:.
\end{equation}
Furthermore, 
\begin{equation}\label{eq-chofrev}
\Bigl[ \A \in \NAmb, \ \  S(\A) \in \Lip \Bigr] \ \implies \
\Bigl[ \A \in \Twin \Bigr]\:. 
\end{equation}
The twin property can be checked in polynomial time, hence Theorem
\ref{th-chof} follows from the above implications. 


\section{Hierarchy of Series}\label{sec:Hier}
The examples in this section illustrate the classes of series on which
we work.
\begin{equation*}
\Seq
\varsubsetneq
\diagramm{(\NAmb\cap\FUSeq)}{\ref{sec:nS/NA/FUS}}
\begin{array}{l}
  \hspace*{.1cm}\tourne{\varsubsetneq}{40}\\
  \tourne{\varsubsetneq}{-30}
\end{array}
\hspace*{.3cm}
\begin{array}{c}
 \diagramm{\FUSeq}{\ref{sec:FUS/nNA}}\\[.6cm]
 \diagramm{\NAmb}{\ref{sec:NA/nFUS}}
\end{array}
\begin{array}{l}
  \tourne{\varsubsetneq}{-40}\\[.4cm]
  \hspace*{.1cm}\tourne{\varsubsetneq}{30}
\end{array}
\hspace*{.3cm}
\diagramm{\FAmb}{\ref{sec:nNA/nFUS/FA}}
\varsubsetneq
\diagramm{\Rat}{\ref{sec:nFA/R}}
\varsubsetneq
\diagramm{\Ser}{\ref{sec:nR}}
\end{equation*}

\subsection{A Series in $\overline{\Seq}\cap \NAmb\cap\FUSeq$}\label{sec:nS/NA/FUS}

An example over a one-letter alphabet is provided in
Figure~\ref{fig:nS/NA/FUS}. The recognized series is
\begin{equation*}
\coef{S}{a^n}=\begin{cases}
0 & \text{ if }n\text{ is odd,}\\
n & \text{ if }n\text{ is even.}
              \end{cases}
\end{equation*}

\begin{figure}[htpb]
\centering
\mbox{\SmallPicture\VCDraw{
\begin{VCPicture}{(-2,-1)(13,1)}
\State[]{(0,0)}{A} \State[]{(3,0)}{B}
\State[]{(8,0)}{C} \State[]{(11,0)}{D}
\Initial{A} \Initial{C}
\Final{B} \Final[s]{C}
\ArcL{A}{B}{a|0}
\ArcL{B}{A}{a|0}
\ArcL{C}{D}{a|1}
\ArcL{D}{C}{a|1}
\end{VCPicture}}}
\caption{$\overline{\Seq}\cap \NAmb\cap \FUSeq$}
\label{fig:nS/NA/FUS}
\end{figure}

The series is not Lipschitz, since
$|\coef{S}{a^{n+1}}-\coef{S}{a^n}|\geq n$, and consequently the series
cannot be sequential (see~(\ref{eq-chof})). It is clear that it is an
unambiguous series (the only successful path of label $a^n$ is the
right or left one depending on the parity of $n$) and a sum of
sequential series.
In fact, any max-plus rational series over a
one-letter alphabet is unambiguous and a sum of sequential
series~\cite{KB94,Moller}.

\begin{figure}[htbp]
\centering
  \SmallPicture\VCDraw{
\begin{VCPicture}{(-1,3)(5,3)}
  \State[]{(0,4)}{A} \State[]{(4,4)}{B} \Initial{A} \Initial{B}
  \Final[w]{A} \Final[w]{B} \LoopN[0.5]{A}{a|1, b|0}
  \LoopN[0.5]{B}{a|0, b|1}
\end{VCPicture}}
\hspace*{3cm}\scalebox{.6}{\begin{picture}(0,0)%
\includegraphics{FUS-nNA.pstex}%
\end{picture}%
\setlength{\unitlength}{4144sp}%
\begingroup\makeatletter\ifx\SetFigFont\undefined%
\gdef\SetFigFont#1#2#3#4#5{%
  \reset@font\fontsize{#1}{#2pt}%
  \fontfamily{#3}\fontseries{#4}\fontshape{#5}%
  \selectfont}%
\fi\endgroup%
\begin{picture}(1124,1484)(3849,-4583)
\end{picture}
}
\caption{$\FUSeq\cap\overline{\NAmb}$}
\label{fig:FUS/nNA}
\end{figure}

\subsection{A Series in $\FUSeq\cap\overline{\NAmb}$}\label{sec:FUS/nNA}

The series $\coef{S}{u}=|u|_a\plus |u|_b$ over the alphabet
$\{a, b\}$ is a sum of two
sequential series: the heap automaton of
Figure~\ref{fig:FUS/nNA} recognizes this series.

Assume that $S$ is unambiguous. The series $S$ is 1-Lipschitz. So it
has to be sequential, see~(\ref{eq-chofrev})
and~(\ref{eq-chof}). Consequently, there exist series $S_1$,\ldots
$S_k$ such that:
$$\forall u\in\ab^*,\,\exists i,\,\exists\lambda_u\in\Rmax\quad
u^{-1}S=\lambda_u\fois S_i.$$

By the pigeon-hole principle, there must exist $i\in\{1,\ldots k\}$
and two integers $m<n$ such that
$$\exists \lambda_n, \lambda_m\quad (a^n)^{-1}S=\lambda_n\fois
S_i,\quad\quad (a^m)^{-1}S=\lambda_m\fois S_i.$$

Consequently, we have
$$\coef{(a^n)^{-1}S}{b^{m+1}}-\coef{(a^n)^{-1}S}{\vide}=
\coef{(a^m)^{-1}S}{b^{m+1}}-\coef{(a^m)^{-1}S}{\vide}.$$

However
\begin{equation*}
  \begin{split}
  \coef{(a^n)^{-1}S}{b^{m+1}}-\coef{(a^n)^{-1}S}{\vide} &
  = \coef{S}{a^nb^{m+1}}-\coef{S}{a^n} = n-n=0\\
  \coef{(a^m)^{-1}S}{b^{m+1}}-\coef{(a^m)^{-1}S}{\vide} &
  = \coef{S}{a^mb^{m+1}}-\coef{S}{a^m} = m+1-m=1.
  \end{split}
\end{equation*}

This is a contradiction, consequently $S$ is not sequential and thus
cannot be an unambiguous series.

\begin{figure}
\centering
\mbox{
\subfigure[]{
\SmallPicture\VCDraw{
\begin{VCPicture}{(-1,2)(5,2)}
\State[]{(0,3)}{A} \State[]{(4,3)}{B}
\Initial{A} \Initial[e]{B}
\Final[w]{A}
\LoopN[.5]{A}{a|1, b|0} \LoopN[.5]{B}{a|1, b|1}
\ArcL{A}{B}{a|1}
\ArcL{B}{A}{a|1}
\end{VCPicture}}
\quad\quad\quad
\scalebox{.5}{\begin{picture}(0,0)%
\includegraphics{NA-nFUS2.pstex}%
\end{picture}%
\setlength{\unitlength}{4144sp}%
\begingroup\makeatletter\ifx\SetFigFont\undefined%
\gdef\SetFigFont#1#2#3#4#5{%
  \reset@font\fontsize{#1}{#2pt}%
  \fontfamily{#3}\fontseries{#4}\fontshape{#5}%
  \selectfont}%
\fi\endgroup%
\begin{picture}(1124,2114)(3849,-4583)
\end{picture}
}}
\quad\quad\quad\subfigure[]{
\SmallPicture\VCDraw{
\begin{VCPicture}{(-1,4)(6,4)}
\State[]{(0,5)}{A} \State[]{(4,5)}{B}
\Initial[e]{B}
\Final[w]{A}
\LoopN[.5]{A}{b|0} \LoopN[.5]{B}{a|1, b|1}
\ArcL{B}{A}{a|1}
\end{VCPicture}}}
}
\caption{$\NAmb\cap\overline{\FUSeq}$}
\label{fig:NA/nFUS0}
\end{figure}

\subsection{Series in $\NAmb\cap \overline{\FUSeq}$}\label{sec:NA/nFUS}
$a)$~The first example is the series $S$ given by the
heap automaton of Figure~\ref{fig:NA/nFUS0}~(a), or equivalently
by the automaton of Figure~\ref{fig:NA/nFUS0}~(b).

\medskip
Consider the series $\tilde{S}$ defined by $\coef{\tilde{S}}{w} =
\coef{S}{w}-|w|$. An automaton recognizing $\tilde{S}$ can clearly be
obtained from an automaton recognizing $S$ by removing 1 from each
output weight. Hence $S$ and $\tilde{S}$ are both sum of sequential
series or none of them is.

The series $\tilde{S}$ is recognized by the automaton of
Figure~\ref{fig:NA/nFUS2}. Suppose that $\tilde{S}=S_1\plus
S_2\plus\cdots \plus S_k$, where $k\in\N$ and the $S_i$ are sequential
series.

Since the $S_i$ are sequential series, they are Lipschitz. Let $N$ be
the maximal Lipschitz coefficient of the $S_i$. Let $(N_i)_{i\geq
  0}$ be a sequence of integers such that
$$N_0>N,\quad N(N_{k-1}+1)<N_k-N_{k-1}\text{ for all }k\geq 1.$$

The coefficient of $ab^{N_k}$ in $\tilde{S}$ is $-N_k$, and it
comes, for instance, from $S_1$. The coefficient of
$ab^{N_k}ab^{N_{k-1}}$ is $-N_{k-1}$. We have:
$$d(ab^{N_k}, ab^{N_k}ab^{N_{k-1}})=N_{k-1}+1\,\text{ and
}\,|\coef{\tilde{S}}{ab^{N_k}}-\coef{\tilde{S}}{ab^{N_k}ab^{N_{k-1}}}|=N_k-N_{k-1}.$$

The coefficient of $ab^{N_k}ab^{N_{k-1}}$ in $\tilde{S}$ does not come from
$S_1$, since
\begin{equation*}
\begin{split}
|\coef{S_1}{ab^{N_k}}- & \coef{S_1}{ab^{N_k}ab^{N_{k-1}}}| \leq
N(N_{k-1}+1)\\
 & < N_k-N_{k-1} =
|\coef{S_1}{ab^{N_k}}-\coef{\tilde{S}}{ab^{N_k}ab^{N_{k-1}}}|.
\end{split}
\end{equation*}

\begin{figure}
\centering
\mbox{\SmallPicture\VCDraw{
\begin{VCPicture}{(-2,2)(4,6)}
\State[]{(0,3)}{A} \Initial[w]{A}
\State[]{(3,3)}{B} \Final[e]{B}
\EdgeL{A}{B}{a|0}
\LoopN[.5]{A}{a|0, b|0}
\LoopN[.5]{B}{b|-1}
\end{VCPicture}}}
  \caption{$\NAmb\cap\overline{\FUSeq}$}
  \label{fig:NA/nFUS2}
\end{figure}

In the same way, we prove that any two words of the set
$$\{ab^{N_k}, ab^{N_k}ab^{N_{k-1}},\ldots ,ab^{N_k}ab^{N_{k-1}}\cdots
ab^{N_0}\}$$ cannot be recognized by the same $S_i$. But this set has
cardinality $k+1$ and thus there is a contradiction.

\bigskip
$b)$~The second example is the series given by the automaton of
Figure~\ref{fig:NA/nFUS1}.
The series recognized by this automaton is:
$$\coef{S}{a^{m_1}b^{n_1}\cdots
  a^{m_p}b^{n_p}}=\sum\limits_{m_i \text{even}}m_i,$$
where $m_1\in\N$,
$m_{k+1}\in\N-\{0\}$, $n_k\in\N-\{0\}$ for $1\leq k\leq p-1$, and
$n_p\in\N$.
The automaton is clearly unambiguous.
Furthermore, it 
is not a finite sum of
sequential series. To simplify notations, let us prove that $S$ is not
the sum of two sequential series. Suppose that $S=S_1\plus S_2$,
with $S_1, S_2\in\Seq$.

The series $S_i$, $i\in\{1, 2\}$, are sequential, so they are
Lipschitz by~(\ref{eq-chof}). Let $N$ be such that $S_i$, $i\in\{1,2\}$, are
$N$-Lipschitz.
Let us consider words of the form $a^rb^na^s$, with $n>0$. We discuss
on the parity of $r$ and $s$.
The coefficient of the word $a^{2p+1}b^na^{2q+1}$ in $S$, which
is equal to 0, comes from one of the $S_i$. For instance
\begin{equation}\label{eq:impair}
\coef{S}{a^{2p+1}b^na^{2q+1}}=0=\coef{S_1}{a^{2p+1}b^na^{2q+1}}.
\end{equation}




Set $q>N$. Since $S_1$ is $N$-Lipschitz and $d(a^{2p+1}b^na^{2q+1},
a^{2p+1}b^na^{2q})=1$, we have
\begin{equation}\label{eq:pair}
\coef{S}{a^{2p+1}b^na^{2q}}=2q=\coef{S_2}{a^{2p+1}b^na^{2q}}.
\end{equation}

Fix $q$ and $n$. Since $S_1$ and $S_2$ are
Lipschitz, there exists an integer $M$ such that:
\begin{equation}\label{eq:DU2}
\forall u,v\in\supp{S_i},\,\, d(u,v)\leq 2n+4q+2\,\Rightarrow\,
|\coef{S_i}{u}-\coef{S_i}{v}|\leq M.
\end{equation}

We have:
\begin{equation*}
d(a^{2p}b^na^{2q}, a^{2p+1}b^na^{2q+1})=2n+4q+2\quad \text{and}\quad
d(a^{2p}b^na^{2q}, a^{2p+1}b^na^{2q})=2n+4q+1.
\end{equation*}
So, by Equation~(\ref{eq:DU2}), we know that:
\begin{itemize}
\item[--] If $a^{2p}b^na^{2q}\in\supp{S_1}$, then
\begin{equation*}
2p+2q=|\coef{S_1}{a^{2p}b^na^{2q}}-\coef{S_1}{a^{2p+1}b^na^{2q+1}}|\leq
M,
\end{equation*}
which is wrong for $p$ large enough.
\item[--] If $a^{2p}b^na^{2q}\in\supp{S_2}$, then
\begin{equation*}
2p=|\coef{S_2}{a^{2p}b^na^{2q}}-\coef{S_2}{a^{2p+1}b^na^{2q}}|\leq M,
\end{equation*}
which is also wrong for $p$ large enough.
\end{itemize}
Consequently, $S$ is not the sum of two sequential series. To extend
 the result to the sum of $m$ sequential series, one has
to consider words of the form $a^{r_1}b^{n_1}a^{r_2}\cdots
a^{r_{m-1}}b^{n_{m-1}}a^{r_m}$.

\begin{figure}
\centering
\mbox{\SmallPicture\VCDraw{
\begin{VCPicture}{(-2,-2)(4,6)}
\State[i]{(0,4)}{A} \Initial{A}
\State[]{(3,4)}{B} \Final[e]{B}
\State[j]{(0,0)}{C} \Initial{C} \Final[w]{C} 
\State[]{(3,0)}{D}
\ArcL{A}{B}{a|0}
\ArcL[.6]{B}{A}{a|0}
\ArcL[.6]{C}{D}{a|1}
\ArcL{D}{C}{a|1}
\VArcR{arcangle=-80}{B}{A}{b|0}
\EdgeL{B}{C}{b|0}
\ArcL{C}{A}{b|0}
\LoopS{C}{b|0}
\end{VCPicture}}}
\caption{$\NAmb\cap \overline{\FUSeq}$}
\label{fig:NA/nFUS1}
\end{figure}

\subsection{Series in $\overline{\NAmb}\cap
 \overline{\FUSeq}\cap\FAmb$}\label{sec:nNA/nFUS/FA}
\begin{figure}[h]
\centering
\mbox{
\subfigure[]{
\SmallPicture\VCDraw{
\begin{VCPicture}{(-.5,3)(5,3)}
\State{(0,3)}{A} \State{(4,3)}{B}
\State{(3,6)}{C}
\Initial{A} \Initial[e]{B} \Initial{C}
\Final[w]{A} \Final{C}
\LoopN[.5]{A}{a|0, b|1, c|0} \LoopN[.75]{B}{a|0, b|1, c|1}
\ArcL{A}{B}{b|1}
\ArcL{B}{A}{b|1}
\LoopN{C}{a|1, b|0, c|0}
\end{VCPicture}}
\hspace*{1.8cm}
{\epsfig{file=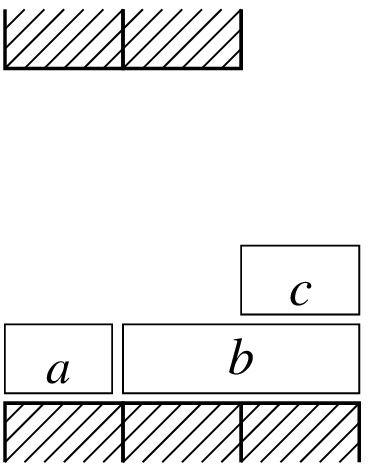, height=2.5cm}}}
\hspace*{1.8cm}\subfigure[]{
\SmallPicture\VCDraw{
\begin{VCPicture}{(-1,3)(7,3)}
\State{(0,3)}{A} \State{(4,3)}{B}
\State{(3,6)}{C}
\Initial{A} \Initial[e]{B} \Initial{C}
\Final[w]{A} \Final{C}
\LoopN[.5]{A}{a|0, c|0} \LoopN[.75]{B}{a|0, b|1, c|1}
\EdgeL{B}{A}{b|1}
\LoopN[.2]{C}{a|1, b|0, c|0}
\end{VCPicture}}}
}
\caption{$\overline{\NAmb}\cap \overline{\FUSeq}\cap\FAmb$}
\label{fig:nNA/nFUS/FA2}
\end{figure}

$a)$~Consider the heap automaton given in
Figure~\ref{fig:nNA/nFUS/FA2}~(a). The corresponding series is at most
two-ambiguous since it is also recognized by the two-ambiguous
automaton of Figure~\ref{fig:nNA/nFUS/FA2}~(b). It cannot be
unambiguous: on $\{a, b\}^*$, since it coincides with the series of
Figure~\ref{fig:FUS/nNA} which is in $\overline{\NAmb}$. It cannot be
a finite sum of sequential series: on $\{b, c\}^*$, it coincides with
the series of Figure~\ref{fig:NA/nFUS0} which is in
$\overline{\FUSeq}$.

$b)$~Another example is provided by the automaton ${\cal A}$ of
Figure~\ref{fig:nNA/nFUS/FA1}.

\begin{figure}
\centering
\mbox{\SmallPicture\VCDraw{
\begin{VCPicture}{(-2,-2)(11,6)}
\State[]{(0,4)}{A} \Initial{A}
\State[]{(3,4)}{B} \Final[e]{B}
\State[]{(0,0)}{C} \Initial{C} \Final[w]{C} 
\State[]{(3,0)}{D}
\ArcL{A}{B}{a|0}
\ArcL[.6]{B}{A}{a|0}
\ArcL[.6]{C}{D}{a|1}
\ArcL{D}{C}{a|1}
\VArcR{arcangle=-80}{B}{A}{b|0}
\EdgeL{B}{C}{b|0}
\ArcL{C}{A}{b|0}
\LoopS{C}{b|0}

\VCPut{(2,0)}{\State[]{(7,4)}{A1} \Initial{A1}
\State[]{(10,4)}{B1} \Final[e]{B1}
\State[]{(7,0)}{C1} \Initial{C1} \Final[w]{C1} 
\State[]{(10,0)}{D1}
\ArcL{A1}{B1}{b|0}
\ArcL[.6]{B1}{A1}{b|0}
\ArcL[.6]{C1}{D1}{b|1}
\ArcL{D1}{C1}{b|1}
\VArcR{arcangle=-80}{B1}{A1}{a|0}
\EdgeL{B1}{C1}{a|0}
\ArcL{C1}{A1}{a|0}
\LoopS{C1}{a|0}
}
\end{VCPicture}}}
\caption{$\overline{\NAmb}\cap \overline{\FUSeq}\cap\FAmb$}
\label{fig:nNA/nFUS/FA1}
\end{figure}

Denote by $S$ the series recognized by
this automaton, by $S_1$ the series recognized by the left part, say
${\cal A}_1$, of the automaton, and by $S_2$ the series
recognized by the right part, say ${\cal A}_2$.

The automaton ${\cal A}_1$ is the one introduced in
Section~\ref{sec:NA/nFUS} and the automaton ${\cal A}_2$ is the same
one after permutation of the $a$'s and $b$'s in the labels. Recall that
${\cal A}_1$ and ${\cal A}_2$ are unambiguous, so $S$ is at most
two-ambiguous.

\medskip

Let us prove that $S$ is not a finite sum of sequential series.
Denote by $L$ the language of words whose blocks of $b$'s have odd
length. Let $u$ be a word of $L$: in ${\cal A}_2$, the $b$-blocks of
$u$ are always read in the upper part of the automaton, so
$\coef{S_2}{u}=0$. Since the coefficient of $u$ in ${\cal A}_1$ is at
least 0, we have $S\odot \1_L=S_1\odot \1_L$.
Suppose that $S$ is a finite sum of sequential series. Then so is
$S\odot \1_L$  and $S_1\odot \1_L$. And this is false since one can choose
an odd $n$ in the proof of Section~\ref{sec:NA/nFUS} for the automaton
of Figure~\ref{fig:NA/nFUS1}.

Let us prove that $S$ is not unambiguous.
Let $M$ be the rational language of words whose $a$-blocks and $b$-blocks have
even lengths and let $u$ be a word of $M$.
In ${\cal A}_1$, the $a$-blocks have to be read in the lower part of
${\cal A}_1$ and so $\coef{S_1}{u}=|u|_a$. In the same way:
$\coef{S_2}{u}=|u|_b$. So we have $S\odot \1_M=S'\odot \1_M$, where $S'$ is
the series recognized by the automaton of Figure~\ref{fig:FUS/nNA}.
Consequently, if $S$ is unambiguous, so is $S'\odot \1_M$. We now apply
the arguments of Section~\ref{sec:FUS/nNA} to show that $S'\odot \1_M$ is
not unambiguous. 

\medskip

$c)$ Besides, Weber has given examples of series which are
$k$-ambiguous and not $(k-1)$-ambiguous~\cite[Theorem~4.2]{w94}.

\subsection{Series in $\overline{\FAmb}\cap\Rat$}\label{sec:nFA/R}
Consider the series $S$ recognized by the automaton of
Figure~\ref{fig:nFA/R1}.
Assume that $S$ is finitely ambiguous. Using the result of
Corollary~\ref{cor-famb2amb} below,  $S$ is recognized by a finite
union of unambiguous automata with the same support, say ${\cal A}_1,
\ldots , {\cal A}_k$.

\begin{figure}
\centering
\mbox{
\subfigure[]{
\SmallPicture\VCDraw{
\begin{VCPicture}{(-1,5)(4,5)}
\State[]{(0,6)}{A} \Initial[w]{A} \Final[w]{A}
\State[]{(4,6)}{B} \Initial[e]{B} \Final[e]{B}
\ArcL[.5]{A}{B}{c|1}
\ArcL[.5]{B}{A}{c|1}
\LoopN[.5]{A}{a|1, b|0, c|1}
\LoopN[.5]{B}{a|0, b|1, c|1}
\end{VCPicture}}}\hspace*{2cm}
\subfigure[]{
\scalebox{.5}{\begin{picture}(0,0)%
\includegraphics{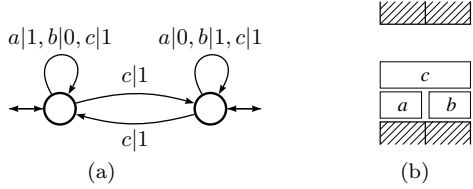}%
\end{picture}%
\setlength{\unitlength}{4144sp}%
\begingroup\makeatletter\ifx\SetFigFont\undefined%
\gdef\SetFigFont#1#2#3#4#5{%
  \reset@font\fontsize{#1}{#2pt}%
  \fontfamily{#3}\fontseries{#4}\fontshape{#5}%
  \selectfont}%
\fi\endgroup%
\begin{picture}(1124,1754)(3849,-4583)
\end{picture}
}}
}
\caption{$\overline{\FAmb}\cap \Rat$}
\label{fig:nFA/R1}
\end{figure}

Denote by $S_i$ the series recognized by ${\cal A}_i$, for $1\leq
i\leq k$, and by $n$ the maximal dimension of an automaton ${\cal A}_i$.
Observe that $\supp{S}=\ab^*$. Since all the $S_i$ have the same
support, we have $\supp{S_i}=\ab^*$.

Now, consider the word $w_0=\bigl(a^nb^nc\bigr)^k$. For any $i$, there
is a single successful path labelled by $w_0$ in ${\cal A}_i$. Note that
a path of length $n$ contains necessarily a circuit.

So, each automaton ${\cal A}_i$ contains a path of the form:


\begin{center}
\mbox{$\pi_i:$\quad
  \SmallPicture\FixVCScale{.43}\VCDraw{
\begin{VCPicture}{(-1,-3)(24,3)}
\SmallState
  \State[]{(0,0)}{A}  \Initial[w]{A}
  \State[]{(3,0)}{B}
  \State[]{(6,0)}{C}
  \State[]{(9,0)}{D}
  \State[]{(12,0)}{E}
  \State[]{(14,0)}{F}
  \State[]{(17,0)}{G}
  \State[]{(20,0)}{H}
  \State[]{(23,0)}{I}  \Final[e]{I}

\EdgeR{A}{B}{a\cdots a}
\EdgeR{B}{C}{a\cdots a}
\EdgeR{C}{D}{b\cdots b}
\EdgeR{D}{E}{b\cdots b}
\EdgeR{E}{F}{c}
\EdgeR{F}{G}{a\cdots a}
\EdgeR{G}{H}{a\cdots a}
\LoopN[.5]{B}{a\cdots a}
\LoopN[.5]{D}{b\cdots b}
\LoopN[.5]{G}{a\cdots a}

\ChgEdgeLineStyle{dashed}
\EdgeR{H}{I}{}
\end{VCPicture}}}
\end{center}


For every $j\in \{1,\ldots ,k\}$, we choose in the subpath labelled by the
$j$-th factor $a^n$ (resp. $b^n$) a circuit that is called the $j$-th
$a$-loop (resp. the $j$-th $b$-loop).

The coefficient of a word in $S$ is less than or equal to its
lentgh, it is thus the same for its coefficients in the
$S_i$. Consequently, the mean weights of the loops of $\pi_i$ are less
than or equal to 1. Denote by $\avw{\pi_i}{a}{j}$ the mean weight of
the $j$-th $a$-loop in the path $\pi_i$, and define
$\avw{\pi_i}{b}{j}$ similarly.

Set $j\in\{1,\ldots ,k\}$. For $\lambda\in\N-\{0\}$, consider the
word $$w_{\lambda}= (a^nb^nc)\cdots(a^nb^nc)
     \underbrace{(a^{n+\lambda n!}b^{n+\lambda n!}c)}_{j\text{-th
     block}}
     (a^nb^nc)\cdots(a^nb^nc)$$

This word can be read on each path $\pi_i$ by turning into  the $j$-th
$a$- and $b$-loops, whose lengths are less than or equal to $n$ and so
divide $n!$.

Let $i\in\{1,\ldots, k\}$ be such that $\coef{S}{w_0}=\coef{S_i}{w_0}$.
We have $\coef{S}{w_{\lambda}}-\coef{S}{w_0}= \lambda n!$, and so
$\coef{S_i}{w_{\lambda}}-\coef{S_i}{w_0}\leq \lambda n!$.
But $\coef{S_i}{w_{\lambda}}-\coef{S_i}{w_0}=(\avw{\pi_i}{a}{j}
+\avw{\pi_i}{b}{j}) \lambda n!$, consequently
\begin{equation}\label{eq:sumavg}
\avw{\pi_i}{a}{j} +\avw{\pi_i}{b}{j}\leq 1.
\end{equation}

Consider any $u$ in $\{01, 10\}^k$. For all $p\in\N-\{0\}$, let us
define the word
$$v_p(u)= (a^{n+\lambda_1 n!}b^{n+\mu_1 n!}c)\cdots
     (a^{n+\lambda_j n!}b^{n+\mu_j n!}c)\cdots
     (a^{n+\lambda_k n!}b^{n+\mu_k n!}c),$$

where $(\lambda_j,\mu_j)=(p,0)$ if $(u_{2j-1}, u_{2j})=(1,0)$ (we say
then that the dominant $j$-th loop is the $j$-th $a$-loop) and
$(\lambda_j,\mu_j)=(0,p)$ otherwise (the dominant $j$-th loop is the
$j$-th $b$-loop), for any $j\in\{1,\ldots ,k\}$.

By the pigeon-hole principle, for some $i$, there are infinitely many
words of the form $v_p(u)$ such that
$\coef{S}{v_p(u)}=\coef{S_i}{v_p(u)}=k+nk+kpn!$. Such words are read
on the path $\pi_i$. The
dominant $j$-th loop in $\pi_i$ has then necessarily mean weight 1,
and by Equation~(\ref{eq:sumavg}), the non-dominant $j$-th loop in
$\pi_i$ has mean weight less than or equal to 0.

Consequently, we have built an injection from the language $\{01, 10\}^k$
into the set of paths $\{\pi_i\}$. But the language has cardinality $2^k$
and the set of paths has cardinality $k$. So we have a contradiction.

\subsection{Rational and Non-Rational Series ($\overline{\Rat}$)}\label{sec:nR}
A max-plus series is non-rational as soon as its support is a non-rational language.
Here, we present a less trivial example of non-rational max-plus
series.

In this paragraph, it is necessary to distinguish between
$\Rmin$ and $\Rmax$: for $R=\Rat$ or $\NAmb$, we use the
respective notations $\Rmin R$, $\Rmax R$.
If $S\in\series{\Rmax}{\ab^*}$, we identify $S$ with
  $\tilde{S}\in\series{\Rmin}{\ab^*}$ such that
  $\coef{\tilde{S}}{w}=\coef{S}{w}$ if $w\in\supp{S}$ and
  $\coef{\tilde{S}}{w}=+\infty$ if $\coef{S}{w}=-\infty$.

Clearly, we have 
$$\Rmax\NAmb=\Rmin\NAmb=\NAmb .$$

On the other hand, it is easy to find
$S\in\Rmin\FUSeq\cap\Rmin\overline{\NAmb}$ such that
$S\not\in\Rmax\Rat$.

Consider for instance the series $S=\min(|w|_a, |w|_b)$ (recognized by
the automaton of Figure~\ref{fig:FUS/nNA} seen as a min-plus
automaton).
Let us prove that $S$ does not belong to $\Rmax\Rat$.
If it does: let
$S_1$, \ldots $S_n$ be a minimal generating family of
$\langle u^{-1}S, u\in \ab^*\rangle$ (see \S\ref{sse-as}), we have:
$\forall u\in\ab^*,\,\exists \lambda_1^{(u)},\ldots
\lambda_n^{(u)},\,u^{-1}S= \bigplus_{i} \lambda_i^{(u)}\fois S_i$.
The restrictions of the quotients of $S$ to $b^*$ are bounded, hence
so are the restrictions of the $S_i$.
Let $k_i$ be such that: $\coef{S_i}{b^{k_i}}=\max_{k}
\coef{S_i}{b^k}$. It follows that for any word $u$: $\max_{k}
\coef{u^{-1}S}{b^k}=\max_{k_i}
\coef{u^{-1}S}{b^{k_i}}$. Consider $k>\max_{i} k_i$. Then
arises a contradiction:
$$\max_{l}
  \coef{(a^k)^{-1}S}{b^l}=k>\max_{k_i}
  \coef{(a^k)^{-1}S}{b^{k_i}}=\max_i k_i.$$

\subsection{Ambiguity vs. sequentiality and Ambiguity vs. Lipschitz}
Here are some examples of series that are in several classes described
in Section~\ref{sec:Hier}:

\begin{center}
\begin{tabular}{|c|c|c|c|}
\hline
{\Large \strut} & $\NAmb$ & $\FAmb\cap \overline{\NAmb}$ & $\overline{\FAmb}$\\ \hline
$\Seq$
& \begin{tabular}[c]{c}
    {\epsfig{file=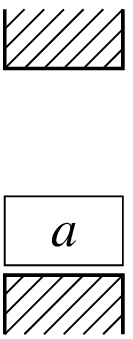, height=1.5cm}}
  \end{tabular}
& impossible
& impossible \\ \hline
  \begin{tabular}{c}
    $\FUSeq$\\
    $\cap$\\
    $\overline{\Seq}$
  \end{tabular}
& \mbox{\SmallPicture\VCDraw{
  \begin{VCPicture}{(-1.5,-4)(4.5,1)}
    \State[]{(0,0)}{A} \State[]{(3,0)}{B} \State[]{(0,-2.5)}{C}
    \State[]{(3,-2.5)}{D} \Initial{A} \Initial{C} \Final{B}
    \Final[s]{C} \ArcL{A}{B}{a|0} \ArcL{B}{A}{a|0} \ArcL{C}{D}{a|1}
    \ArcL{D}{C}{a|1}
  \end{VCPicture}}}
& \begin{tabular}{c}
    {\epsfig{file=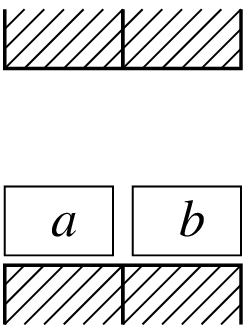, height=1.5cm}}\\
    Sec~\ref{sec:FUS/nNA}
  \end{tabular}
& impossible \\ \hline
$\overline{\FUSeq}$
& \begin{tabular}[c]{c}
    {\epsfig{file=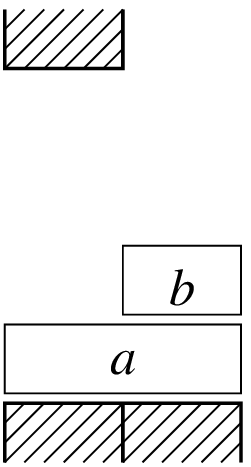, height=2.5cm}}\\
    Sec~\ref{sec:NA/nFUS}
  \end{tabular}
& \begin{tabular}[c]{c}
    {\epsfig{file=nL-FA.eps, height=2.5cm}}\\
    Sec~\ref{sec:nNA/nFUS/FA}
  \end{tabular}
& \begin{tabular}[c]{c}
  {\epsfig{file=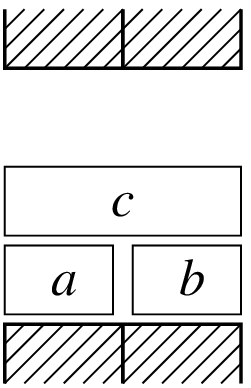, height=2.5cm}}\\
    Sec~\ref{sec:nFA/R}
  \end{tabular}
\\ \hline
\end{tabular}

\begin{tabular}{|c|c|c|c|}
\hline
{\Large \strut} & $\NAmb$ & $\FAmb\cap \overline{\NAmb}$ & $\overline{\FAmb}$\\
 \hline
$\Lip$
& \begin{tabular}[c]{c}
    {\epsfig{file=S-NA.eps, height=1.5cm}}
  \end{tabular}
& \begin{tabular}[c]{c}
    {\epsfig{file=FUS-nNA.eps, height=1.5cm}}\\
    Sec~\ref{sec:FUS/nNA}
  \end{tabular}
& \begin{tabular}[c]{c}
    {\epsfig{file=nFA-R1.eps, height=2cm}}\\
    Sec~\ref{sec:nFA/R}
  \end{tabular}
\\ \hline
$\overline{\Lip}$
& \begin{tabular}[c]{c}
    {\epsfig{file=NA-nFUS2.eps, height=2cm}}\\
    Sec~\ref{sec:NA/nFUS}
  \end{tabular}
& \begin{tabular}[c]{c}
    {\epsfig{file=nL-FA.eps, height=2cm}}\\
    Sec~\ref{sec:nNA/nFUS/FA}
  \end{tabular}
& \begin{tabular}[c]{c}
    {\epsfig{file=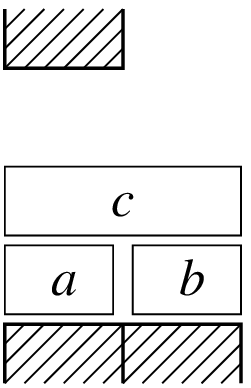, height=2cm}}
  \end{tabular}\\ \hline
\end{tabular}
\end{center}


\section{From Finitely Ambiguous to Union of Unambiguous}
Weber~\cite{w94} has proved that a finitely ambiguous $\Nmax$-automaton
can be turned into an union of unambiguous ones.
We present a completely different and simpler proof
that holds in any semiring, in particular $\Rmax$.

In this section, we work on the structure of the automata. So
we consider simply Boolean automata.

Below, given a set $S$, we identify the vectors of $\B^S$ with the subsets of
$S$, \ie $x\in\B^S$ is identified with $\{i\in S\mid x_i=\1\}$.

Let $\A=(\alpha,\mu,\beta)$ be a trim
automaton. The {\em past} of a state $p$ is the set of words that
label a path from some initial state to $p$. The {\em future} of $p$
is the set of words that label a path from $p$ to some final state. We
write: $$\Past[\A]{p}=\{w\in\ab^*\mid (\alpha\mu(w))_p=\un\},\qquad 
\Fut[\A]{p}=\{w\in\ab^*\mid (\mu(w)\beta)_p=\un\}.$$  

Let $\A=(\alpha,\mu:\ab^*\rightarrow\B^{Q\times Q},\beta)$ be an
automaton. Let us recall the usual determinization procedure of $\A$
{\it via} the subset construction. Let $R$ be the least subset of
$\B^Q$ inductively defined by: $$\alpha\in R,\quad X\in R\Rightarrow
\forall a\in\ab, X\mu(a)\in R.$$

Let $\D=\D(\cA)=(J,\nu:\ab^*\rightarrow\B^{R\times R},U)$ be the {\em determinized automaton} of~$\A$ defined by:
$$J=\{\alpha\},\qquad U=\{P\in R\mid P\beta=\un\},
\qquad\nu(a)_{P,P'}=\un \Longleftrightarrow P'=P\mu(a).$$

\begin{lem}\label{prop-detprop}
$i)$ Let $\A$ be an automaton and $\D$ its determinized automaton.
Then for each state $P$ of $D$,
$$\Past[\D]{P}\subseteq\bigcap_{p\in P}\Past[\A]{p},\quad\text{ and }\quad\Fut[\D]{P}=\bigcup_{p\in P}\Fut[\A]{p}.$$
$ii)$ Let $\A$ and $\Bc$ be two automata and $\A\odot\Bc$ their tensor
product ({\it cf.\/} \S\ref{sse-as}), then,
for all state~$(p,q)$ of~$\A\odot\Bc$,
$$\Past[\A\odot\Bc]{p,q}=\Past[\A]{p}\cap\Past[\Bc]{q},
\qquad
 \Fut[\A\odot\Bc]{p,q}=\Fut[\A]{p}\cap\Fut[\Bc]{q}.$$
\end{lem}

The constructions and results given in Propositions~\ref{prop-past} and~\ref{prop-US}
are inspired by Sch\"utzenberger~\cite{Sc76}. 
They have been explicitely stated by Sakarovitch in~\cite{Sa98}.

Let $\A$ be an automaton and $\D$ its determinized automaton.
The trim part of the product $\A\odot\D$ is called the {\em Sch\"utzenberger
  covering $\Sc$ of $\A$}.

\begin{figure}[t]
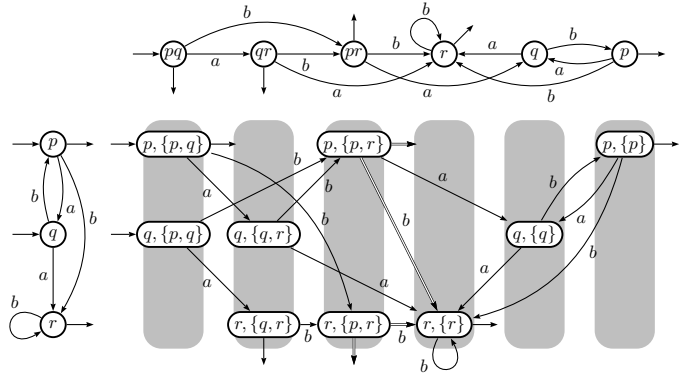

\begin{center}
\FixVCScale{.4}
\VCDraw{%
\begin{VCPicture}{(-1,-2)(19,10)}
\psframe*[linecolor=lightgray,framearc=.6](2,-.8)(4,6.8)
\psframe*[linecolor=lightgray,framearc=.6](5,-.8)(7,6.8)
\psframe*[linecolor=lightgray,framearc=.6](8,-.8)(10,6.8)
\psframe*[linecolor=lightgray,framearc=.6](11,-.8)(13,6.8)
\psframe*[linecolor=lightgray,framearc=.6](14,-.8)(16,6.8)
\psframe*[linecolor=lightgray,framearc=.6](17,-.8)(19,6.8)

\State[p]{(-1,6)}{X}
\State[q]{(-1,3)}{Y}
\State[r]{(-1,0)}{Z}
\State[pq]{(3,9)}{A}
\State[qr]{(6,9)}{B}
\State[pr]{(9,9)}{C}
\State[r]{(12,9)}{D}
\State[q]{(15,9)}{E}
\State[p]{(18,9)}{F}
\StateVar[p,\{p,q\}]{(3,6)}{AX}
\StateVar[q,\{p,q\}]{(3,3)}{AY}
\StateVar[q,\{q,r\}]{(6,3)}{BY}
\StateVar[r,\{q,r\}]{(6,0)}{BZ}
\StateVar[p,\{p,r\}]{(9,6)}{CX}
\StateVar[r,\{p,r\}]{(9,0)}{CZ}
\StateVar[r,\{r\}]{(12,0)}{DZ}
\StateVar[q,\{q\}]{(15,3)}{EY}
\StateVar[p,\{p\}]{(18,6)}{FX}
\Initial{X}
\Initial{Y}
\Final{X}
\Final{Z}
\ArcL[.8]{X}{Y}{a}
\LArcL{X}{Z}{b}
\EdgeR{Y}{Z}{a}
\ArcL{Y}{X}{b}
\LoopW{Z}{b}
\Initial{A}
\Final[s]{A}
\Final[s]{B}
\Final[n]{C}
\Final[ne]{D}
\Final{F}
\ArcL[.8]{F}{E}{a}
\LArcL{F}{D}{b}
\EdgeR{E}{D}{a}
\ArcL{E}{F}{b}
\LoopL[.7]{135}{D}{b}
\EdgeR{A}{B}{a}
\LArcL{A}{C}{b}
\LArcR{B}{D}{a}
\EdgeR{B}{C}{b}
\LArcR{C}{E}{a}
\EdgeL{C}{D}{b}
\Initial{AX}
\Initial{AY}
\Final{AX}
\Final[s]{BZ}
\Final{DZ}
\Final{FX}
\ArcL[.8]{FX}{EY}{a}
\LArcL{FX}{DZ}{b}
\EdgeR{EY}{DZ}{a}
\ArcL{EY}{FX}{b}
\LoopVarS{DZ}{b}
\EdgeR{AX}{BY}{a}
\EdgeR{AY}{BZ}{a}
\LArcL[.6]{AX}{CZ}{b}
\EdgeL[.8]{AY}{CX}{b}
\EdgeL[.7]{BY}{DZ}{a}
\EdgeR[.7]{BY}{CX}{b}
\EdgeR{BZ}{CZ}{b}
\EdgeL{CX}{EY}{a}
\FixEdgeLineDouble{0.6}{1.2}            
\EdgeLineDouble
\Final{CX}
\Final[s]{CZ}
\EdgeL{CX}{DZ}{b}
\EdgeR{CZ}{DZ}{b}
\end{VCPicture}}
\end{center}
\caption{{\small$\A$ (left), the determinized automaton (top) and the
    Sch\"utzenberger covering}}\label{figschut1}
\end{figure}
\begin{prop}\label{prop-past}\ 
Let $\A=(\alpha, \mu, \beta)$ be a trim automaton, $\D$ its determinized
automaton and $\Sc$ its Sch\"utzenberger covering.
\\$i)$
The states of $\Sc$ are exactly the pairs $(p,P)$, where $P$ is a
state of $\D$ and $p\in P$. We call the set $\{(p,P)\mid p\in P\}$ of
states of $\Sc$ a {\em column} (in gray on Figure~\ref{figschut1}).
\\$ii)$
The canonical surjection $\psi$ from the transitions of~$\Sc$ onto the
transitions of~$\A$ induces a one-to-one mapping between the
successful paths of $\Sc$ and~$\A$.
\\$iii)$
Let~$P$ be a state of $\D$. Then, for every $p$ in $P$,
$$\Past[\Sc]{p,P}=\Past[\D]{P},\qquad  \Fut[\Sc]{p,P}=\Fut{p}.$$
Thus, all the states of a given column have the same past.
\end{prop}

\begin{pf}
$i)$ A state $(p,P)$ of $\Sc$ is initial if and only if $p$ is initial
in $\A$ (\ie $p\in \alpha$) and $P$ is initial in $\D$ (\ie $P=\{\alpha\}$).
Now, let $(p,P)$ be a state of $\Sc$ such that $p\in P$ and $(q,Q)$ a
successor of $(p,P)$ by $a$.
Then, there exist two transitions:
$$\chemin{p\flech{a}q}{\Ac}\quad\text{and}\quad\chemin{P\flech{a}Q}{\D}.$$
By definition of $\D$, $q$ belongs thus to $Q$.

Conversely, let $P$ be a state of $\D$ and $p$ an element of $P$.
For every $w$ in $\Past[\D]{P}$,
$w$ belongs to $\Past[\A]{p}$ (Lemma~\ref{prop-detprop}). Therefore there is a path in $\Sc$
from an initial state to $(p,P)$. 

$ii)$ Let $\pi$ be a successful path of $\A$, with label $w=w_1w_2\cdots w_n$.
Let $\theta$ be the (unique) successful path with label $w$ in $\D$:
$$\pi=\chemin{\init{p_0}\flech{w_1}p_1\flech{w_2}\ldots\flech{w_n}\final{{p_n}}}{\Ac},\quad
\theta=\chemin{\init{P_0}\flech{w_1}P_1\flech{w_2}\ldots\flech{w_n}\final{{P_n}}}{\D}.$$
There is a path in $\Sc$:
$\pi'=\init{(p_0,P_0)}\flech{w_1}{(p_1,P_1)}\flech{w_2}\ldots\flech{w_n}\final{(p_n,P_n)}$.
\\The function~$\pi\mapsto\pi'$ is obviously one-to-one.

$iii)$ By results from Lemma~\ref{prop-detprop}:
$$
\left.
\begin{array}{c}
\Past[\Sc]{p,P}=\Past[\A]{p}\cap\Past[\D]{P}\\[.2em]
\forall p\in P,\Past[\D]{P}\subseteq\Past[\A]{p}
\end{array}\right\}
\Rightarrow
\forall p\in P,\Past[\Sc]{p,P}=\Past[\D]{P}.$$
$$
\left.
\begin{array}{c}
\Fut[\Sc]{p,P}=\Fut[\A]{p}\cap\Fut[\D]{P}\\[.2em]
\forall p\in P,\Fut{p}\subseteq\Fut[\D]{P}
\end{array}\right\}
\Rightarrow
\forall p\in P,\Fut[\Sc]{p,P}=\Fut{p}.$$
\hspace*{\fill}\qed
\end{pf}

\begin{defn}
In $\Sc$, different transitions with the same label, the same
destination and whose origins belong to the same column are said to be
{\em competing}. Likewise, different final states of the same column
are {\em competing}.  A {\em competing set} is a maximal set of
competing transitions or competing final states.
\end{defn}
Let~$\Uc$ be an automaton obtained from $\Sc$ by removing
all transitions except one in every competing set and 
by turning all final states of a column, except one, into non-final states.
The choice of the transition (or the final state) to keep
in a competing set is arbitrary.

\begin{figure}[htbp]
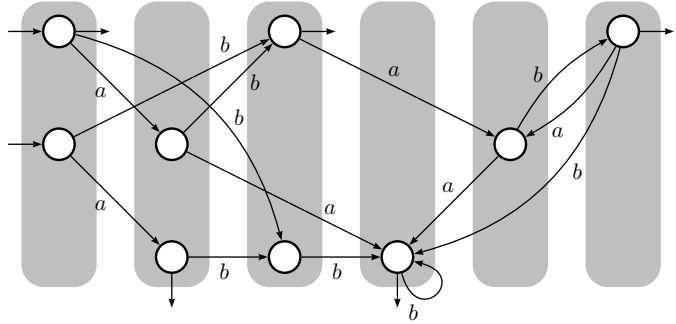

\centering
    \FixVCScale{.5}
\VCDraw{%
\begin{VCPicture}{(1.5,-2.5)(19.5,7)}
  \psframe*[linecolor=lightgray,framearc=.6](2,-.8)(4,6.8)
  \psframe*[linecolor=lightgray,framearc=.6](5,-.8)(7,6.8)
  \psframe*[linecolor=lightgray,framearc=.6](8,-.8)(10,6.8)
  \psframe*[linecolor=lightgray,framearc=.6](11,-.8)(13,6.8)
  \psframe*[linecolor=lightgray,framearc=.6](14,-.8)(16,6.8)
  \psframe*[linecolor=lightgray,framearc=.6](17,-.8)(19,6.8)
  \State{(3,6)}{AX} \State{(3,3)}{AY} \State{(6,3)}{BY}
  \State{(6,0)}{BZ} \State{(9,6)}{CX} \State{(9,0)}{CZ}
  \State{(12,0)}{DZ} \State{(15,3)}{EY} \State{(18,6)}{FX}
  \Initial{AX} \Initial{AY} \Final{AX} \Final[s]{BZ} \Final[s]{DZ}
  \Final{FX} \ArcL[.8]{FX}{EY}{a} \LArcL{FX}{DZ}{b} \EdgeR{EY}{DZ}{a}
  \ArcL{EY}{FX}{b} \LoopSE{DZ}{b} \EdgeR{AX}{BY}{a} \EdgeR{AY}{BZ}{a}
  \LArcL[.6]{AX}{CZ}{b} \EdgeL[.8]{AY}{CX}{b} \EdgeL[.7]{BY}{DZ}{a}
  \EdgeR[.7]{BY}{CX}{b} \EdgeR{BZ}{CZ}{b} \EdgeL{CX}{EY}{a}
  \Final{CX}
  \EdgeR{CZ}{DZ}{b}
\end{VCPicture}}
\caption{An unambiguous automaton equivalent to $\Sc$}
\label{figschutz}
\end{figure}

For instance, the covering $\Sc$ of Figure~\ref{figschut1}
has two competing sets (drawn with double lines);
the first one contains two transitions with label $b$ that arrive
in $(r,\{r\})$, the second one contains the states $(p,\{p,r\})$ and
 $(r,\{p,r\})$ which are both final. The above selection principle gives rise to four
possible automata,
the automaton of Fig. \ref{figschutz} being one of them.

\begin{prop}\label{prop-US}
Let $\Sc$ and $\Uc$ be two automata defined as above. Then,
\\$i)$ $\forall P,\forall p\in P,\,\Past[\Uc]{p,P}=\Past[\Sc]{p,P}.$
\\$ii)$ Futures of states in a column of $\Uc$ are disjoint
and
$$\forall P,\forall p\in P,\,\bigcup_{p\in P}
\Fut[\Uc]{p,P}=\bigcup_{p\in P} \Fut[\Sc]{p,P}\:.$$ 
\\Consequently, the automaton $\Uc$ is unambiguous and equivalent to~$\A$.
\end{prop}

\begin{pf}
$i)$ The proof is by induction on the length of words.
If $(p,P)$ is initial in $\Sc$, it is still initial in $\Uc$.
Let $wa$ be a word of $\Past[\Sc]{p,P}$ and $\pi$ a path labelled
by this word from an initial state to $(p,P)$. We consider the last transition of $\pi$:
$$\chemin{(q,P')\flech{a}(p,P)}{\Sc}.$$
If this transition does not belong to a competing set, it still
appears in $\Uc$ and, by induction, $w\in \Past[\Uc]{q,P'}$,
thus $wa\in \Past[\Uc]{p,P}$.
If this transition belongs to a competing set, there exist $q'\in P'$
and a transition
$$\chemin{(q',P')\flech{a}(p,P)}{\Sc}$$
which still appears in $\Uc$, and by induction,
since $\Past[\Sc]{q,P'}=\Past[\Sc]{q',P'}$, $w\in\Past[\Uc]{q',P'}$,
so $wa\in \Past[\Uc]{p,P}$.

$ii)$ We prove this by induction on the length of words.
If there are several final states in a column of $\Sc$, exactly one remains
in $\Uc$, so there is at most one state whose future contains the
empty word. 
Now let $(p, P)$ and $(p', P')$ be two states in the same column such
that the word $au$ belongs to $\Fut{p}$ and $\Fut{p'}$:
$$\chemin{p\flech{a}q\flech{u}\final{t}}{\A}$$
$$\chemin{p'\flech{a}q'\flech{u}\final{t'}}{\A}$$
Both transitions $p\flech{a}q$ and $p'\flech{a}q'$
correspond to the same transition in $\D$. Thus $q$ and $q'$
belong to the same column and, by induction, $q=q'$.
Since there is no competing set in $\Uc$, $p=p'$.
\\
Obviously $\Fut[\Uc]{p,P}\subseteq \Fut[\Sc]{p,P}$.
If $au$ is in the future of a state $(p_0,P_0)$ of $\Sc$,
there exist a state $(p_1,P_1)$ and a transition
$(p_0,P_0)\flech{a}(p_1,P_1)$, such that $u$ is in
$\Fut[\Sc]{p_1,P_1}$. By induction, there exists $p'_1$ in~$P_1$ such
that $u$ is in~$\Fut[\Uc]{p'_1,P_1}$, and there exists a transition
$(p'_0,P_0)\flech{a}(p'_1,P_1)$, thus $au$ is
in~$\Fut[\Uc]{p'_0,P_0}$.

Let $w$ be a word accepted by $\A$. For any factorization $uv$ of $w$,
there is exactly one column~$P$ of $\Uc$ such that, for every $p$
in~$P$, $u$ is in~$\Past[\Uc]{p,P}$ and there is exactly
one state~$(p,P)$ in this column such that $v$ is in~$\Fut[\Uc]{p,P}$.
This characterizes the only successful path with label $w$ in $\Uc$.
\hspace*{\fill}\qed
\end{pf}

\medskip

We show now how the Sch\"utzenberger covering can be used to
convert a finitely ambiguous automaton~$\A$ into a finite union of
unambiguous automata, each of them recognizing the same language as~$\A$.

\begin{prop}\label{prop-comp}
Let~$\Sc$ be the Sch\"utzenberger covering of a finitely ambiguous
automaton. Then, competing transitions of~$\Sc$ do not belong to
any circuit of~$\Sc$. Thus a path of~$\Sc$ contains at
most one transition of each competing set.
\end{prop}

\begin{pf}
Assume that a competing transition $\tau$ belongs to a circuit:
$$\init{i}\flech{u}(p,P)\underset{\tau}{\flech{a}}(q,Q)\flech{w}(p,P)\underset{\tau}{\flech{a}}(q,Q)\flech{v}\final{t}.$$
Hence, $u(aw)^*$ is a subset of $\Past{p}$.
Let $\tau'$ be another transition that belongs to the same competing set:
$(p',P)\underset{\tau'}{\flech{a}}(q,Q).$
From Lemma~\ref{prop-detprop}, $u(aw)^*$ is a subset of $\Past[\A]{p'}$.
 Thus, for every $n$, for every $k$ in $\{0,\ldots ,n\}$,
there exists a path:
$$\init{i}\flech{u(aw)^k}(p',P)\underset{\tau'}{\flech{a}}\left[(q,Q)\flech{w}(p,P)\underset{\tau}{\flech{a}}(q,Q)\right]^{n-k}\flech{v}\final{t}.$$
Therefore, there are at least $n+1$ successful paths with label $u(wa)^nv$ in $\Sc$,
which is in contradiction with the finite ambiguity of $\Sc$ and $\A$.

If there exists a path of $\Sc$ that contains two competing transitions
$\tau$ and $\tau'$:
$$(p,P)\underset{\tau}{\flech{a}}(q,Q)\flech{w}(p',P)\underset{\tau'}{\flech{a}}(q,Q),$$
then $\tau'$ belongs to a circuit, which is impossible.
\hspace*{\fill}\qed
\end{pf}

Assume that $\A$ is finitely ambiguous. As a consequence of
Proposition~\ref{prop-comp}, for every path in $\Sc$ (and thus for
every path in $\A$), one can compute an unambiguous automaton $\Uc$
that contains this path. Consider the following algorithm.

As they do not belong to any circuit, competing sets of $\Sc$ are partially ordered.
\begin{itemize}
\item[--] Compute $C$, the set of maximal competing sets of $\Sc$ (there
  is no path from any element of $C$ to another competing set).
\item[--] Let $\Sc_1$ and $\Sc_2$ be two copies of $\Sc$. For every
  competing set $X$ in~$C$, let $x$ be an element of $X$; 
\item[--] if $x$ is a transition, remove every transition of
  $X\moins\{x\}$ in $\Sc_1$ and remove $x$ in~$\Sc_2$; 
\item[--] if $x$ is a final state, make every state of $X\moins\{x\}$ in
  $\Sc_1$ non-final and make $x$ in $\Sc_2$ non-final.
\item[--] Apply inductively this algorithm to $\Sc_1$ and $\Sc_2$.
\end{itemize}
The result is a finite set of unambiguous automata.
Each of them recognizes the language of $\A$ and every path of $\Sc$
appears in at least one of these automata. Notice that the cardinality of
this set may be larger than the degree of ambiguity of $\A$. Denote by
$\mathcal{F}$ the automaton obtained by taking the union of the
automata in this set.

\medskip

Assume now that~$\A$ is any automaton with multiplicities over an
idempotent semiring. Since there is a canonical mapping from the
transitions  (resp. initial states, resp. final states) of the
Sch\"utzenberger covering~$\Sc$ onto the transitions (resp. initial
states, resp. final states) of~$\A$, one can decorate every transition
(resp. initial state, resp. final state) of~$\Sc$ with the
corresponding multiplicity in~$\A$. This decoration can be carried out
in the same way on the automaton $\mathcal{F}$.

Obviously, since there is a one-to-one mapping between the successful
paths of~$\A$ and those of $\Sc$, the series realized by~$\Sc$ is
equal to the one realized by~$\A$.

Furthermore, as every path of $\Sc$ appears in $\mathcal{F}$, the automaton
$\mathcal{F}$ realizes the same series as $\A$. Notice that a path of $\Sc$ may
appear several times in $\mathcal{F}$, with no consequence since
the semiring is idempotent.

The construction of $\mathcal{F}$ could be modified in
order to get a one-to-one relation between paths of $\A$ and paths of
$\mathcal{F}$, but then the automata in the union would not have
the same support, which would be less convenient in the sequel.

\begin{cor}\label{cor-famb2amb}
A finitely ambiguous max-plus automaton can be effectively turned into an
equivalent finite union of unambiguous max-plus automata, all with the same support.
\end{cor}

\section{The Decidability Result}
\label{se-cp}

In this section, we show that a series, realized by a finite union of
unambiguous automata having the same support, is unambiguous if and
only if a certain property denoted by ({\bf P}) holds.
Associated with Theorem~\ref{th-chof} and
Corollary~\ref{cor-famb2amb}, this enables to prove
Theorem~\ref{th-carac}, stated at the end of the paper.

\medskip

Consider a finite family of max-plus automata $(\A_i)_{i\in I}$ with
respective dimensions $(Q_i)_{i\in I}$. Set $\A_i=(\alpha\ind{i},
\mu\ind{i}, \beta\ind{i})$. The corresponding product automaton
${\cal P}$ is an
automaton with multiplicities in the product semiring $\Rmax^I$,
defined as follows. 
\\Set $Q=\displaystyle{\prod_{i\in I} Q_i}$ and consider $A,B\in 
(\Rmax^I)^Q, M: \Sigma^* \rightarrow (\Rmax^I)^{Q\times Q}$
with
\begin{align*}
\forall \p{p},\p{q} \in Q, \  &\ A_{\p{p}} =
(\alpha\ind{i}_{p_i})_{i\in I},\\
&\forall a\in \ab, \ 
M(a)_{\p{p},\p{q}}= \begin{cases} (\mu\ind{i}(a)_{p_i,q_i})_{i\in I}
  & \mbox{ if } \forall i, \ \mu\ind{i}(a)_{p_i,q_i}\neq \zero \\ 
(\zero, \dots, \zero) & \mbox{ otherwise }\end{cases}\\
&\ \ B_{\p{p}}=
(\beta\ind{i}_{p_i})_{i\in I}\:.
\end{align*}
A state $\p{q}\in Q$ is {\em initial} if  $\forall i, \
(A_{\p{q}})_i\neq \zero$. A state $\p{q}\in Q$ is {\em final} if  $\forall i, \
(B_{\p{q}})_i\neq \zero$. The trim part of $(A,M,B)$ with respect to
the above definition of initial and final states is the {\em product
automaton} ${\cal P}$. 

Clearly, if the automata $(\cA_i)_{i\in I}$ are unambiguous and all
have the same support, then the product automaton $\cP$ is also
unambiguous and satisfies 
\begin{equation*}
\begin{split}
\forall u\in \ab^*,\forall i\in I, & \ \coef{S({\cal
    P})}{u}_i=\alpha\ind{i}\mu\ind{i}(u)\beta\ind{i}\\
 & \Rightarrow \ \bigoplus_{i\in I} \coef{S({\cal P})}{u}_i= 
\coef{\bigoplus_{i\in I} S(\A_i)}{u}= \bigoplus_{i\in I}
\alpha\ind{i}\mu\ind{i}(u)\beta\ind{i}.
\end{split}
\end{equation*}

\begin{defn}
Let $\theta$ be a simple circuit of $\P$, whose weight is
$(x\ind{i})_{i\in I}$.
The set of {\em victorious coordinates of $\theta$}, denoted by
$\vict\theta$, is the set
of coordinates on which the weight of $\theta$ is maximal, \ie
$\vict\theta=\bigl\{i\in I\mid x\ind{i}=\displaystyle{\max_{j\in I}}\{x\ind{j}\}\bigr\}$.
\end{defn}

This definition is extended in a natural way to a strongly connected
subgraph $C$ of $\P$: the set of {\em victorious coordinates of
  $C$} is the intersection of the sets of victorious coordinates of
the simple circuits of $C$. We also extend the definition to a path
$\pi$ of $\P$: the set of {\em victorious coordinates of
  $\pi$} is the intersection of the sets of victorious coordinates of
the strongly connected subgraphs of $\P$ crossed by $\pi$. 

\medskip

Let us define the `dominance' property ({\bf P}): \\

{\em For each
    successful path $\pi$ of the product automaton $\P$, the set of
    victorious coordinates of~$\pi$ is not empty.} 

\medskip

Obviously, the number of simple circuits is finite.
Hence ({\bf P}) is a decidable property.

Let
$(\A_i=(\alpha\ind{i}\in\Rmax^{Q_i},
\mu\ind{i}:\ab^*\rightarrow\Rmax^{Q_i\x Q_i},
\beta\ind{i}\in\Rmax^{Q_i})
)_{i\in I}$
be a finite family of
unambiguous trim automata, all with the same support, and let $\P$
be the product automaton with set of states $Q\subseteq\Pi_{i\in I}Q_i$.
We assume that $\P$ satisfies the dominance property ({\bf P}).

\noindent
Let $N = |Q|$ and 
$\displaystyle M = \max(\max_{i,a,p,q}\mu\ind{i}(a)_{p,q},\max_{i,p}\beta\ind{i}_p)
-\min(\min_{i,a,p,q}\mu\ind{i}(a)_{p,q},\min_{i,p}\beta\ind{i}_p)$,
where the minima are taken over non-$\zero$ terms. In words, $M$ is
the difference between the largest and the smallest non-initial
weights appearing in the automata.

We use the following notations as shortcuts. 
For $\p{x}=(x\ind{i})_{i\in I}\in\Rmax^I$, set $\vmin{\p{x}}= \min_{i\in I}
\{x\ind{i}\mid x\ind{i}\neq-\oo\}$ and $\vnorm{\p{x}} \,=\,
\p{x}-(\vmin{\p{x}},\dots,\vmin{\p{x}})$.

\medskip

Set $I=\{1,\dots , n\}$. We now define an automaton
$\Ua$ that is shown to be unambiguous and to
realize the series $\bigoplus_{i\in I}S(\A_i)$.

The states of $\Ua$ belong to $\Rmax^n\x Q$.

\paragraph*{Initial states.} All the initial states are defined as
follows.
If $\p{q}=(q\ind{1},\dots, q\ind{n})$ is a tuple such that $q\ind{i}$ is an
initial state of $\A_i$, and if we set
$\boldsymbol{\alpha}=(\alpha_{q\ind{1}}\ind{1},\dots,\alpha_{q\ind{n}}\ind{n}),$
then $(\vnorm{\boldsymbol{\alpha}}, \p{q})$ is an initial state
of $\Ua$ and the weight of the ingoing arc is $\vmin{\boldsymbol{\alpha}}$. 

\paragraph*{States and transitions.} 
If $(\p{z},\p{p})$
is a state of $\Ua$, then for
each transition in $\P$ of type:
$ \p{p}\xrightarrow{a \mid \p{x}}
  \p{q} $
such that
$x\ind{i}\neq-\oo$ for all $i$,
there is a transition in $\Ua$ leaving $\p{p}$, labelled by the letter
$a$, and that we now describe.
Set
$\p{t}=\p{z}+\p{x}$.
Let $V$ be the set of victorious coordinates of the maximal strongly connected
subgraph of $\p{q}$ in $\P$. Since $\P$ satisfies
({\bf P}), the set $V\cap \{t\ind{k}\neq -\infty\}$ is non-empty. 
Let $j\in V$ be such that $t\ind{j}=\min_{k\in V}\{t\ind{k}\mid t\ind{k}\neq-\oo\}$, 
and let $\p{y}\in\Rmax^n$ be defined by:
$$\forall i,\quad
y\ind{i}=
\begin{cases}
  -\oo    & \text{if $t\ind{i}<t\ind{j}-NM$,}\\
  t\ind{i} &  \text{otherwise.}
\end{cases}
$$

Now $(\vnorm{\p{y}},\,\p{q})$
is a state of $\Ua$ and we have the following transition:
\[
  \chemin{
    (\p{z},\, \p{p})
    \fleche{a}{\vmin{\p{y}}}
    (\vnorm{\p{y}},\,\p{q})
  }{\Ua}\;.
\]

\paragraph*{Final states.} All the final states are defined as
    follows. 
If $(\p{z},\p{q})$
is a state of $\Ua$, and if $q\ind{i}$ is a final state of $\A_i$ for
all $i$, then $(\p{z},\p{q})$ is a final state of $\Ua$  
and the weight of the outgoing arc is $
\max_{i\in I}\{z\ind{i}+\beta_{q\ind{i}}\ind{i}\}$.

\begin{lem}\label{prop:finite}
The set of states of $\Ua$ is finite. 
\end{lem}

\begin{pf}
First, given a state
$(\p{z_1},\p{q})$
of $\Ua$,
we show that there are finitely many states of the form $(\p{z_2},\p{q})$
that can be reached from $(\p{z_1}, \p{q})$.

Observe that a path leading from $(\p{z_1},\p{q})$ to $(\p{z_2},\p{q})$ in $\Ua$
corresponds to a circuit leading from $\p{q}$ to $\p{q}$ in $\P$
that can
be fully decomposed into simple circuits belonging to the strongly connected
component of $\p{q}$.
Let $V$ be the set of victorious coordinates of the strongly connected
component of $\p{q}$.  By definition of victorious
coordinates, for all $i\in V$ the value of $z_2\ind{i}-z_1\ind{i}$ is a
constant, that we denote by $x$, and for all $i\not\in V$ one has
$z_2\ind{i}\leq z_1\ind{i} + x$.


Let ${\cal C}$ be the (finite) set of simple circuits of $\cal P$. For
a circuit $\theta\in {\cal C}$, let the weight of the circuit in $\cal
P$ be denoted by $(\wght{\theta}\ind{1},\dots ,
\wght{\theta}\ind{n})$. Set also $\wght{\theta}=\max_{i\leq n}
\wght{\theta}\ind{i}$. 
Now define
\[
\delta = \min_{\theta \in \cal C} \Bigl[\wght{\theta} - \max_{i}
\{\wght{\theta}\ind{i}\mid\wght{\theta}\ind{i} < \wght{\theta} \}
\Bigr] \:.
\] 
By definition, we have $\delta >0$.  
By construction, for $i\not\in V$, either $z_2\ind{i}= z_1\ind{i} +
x$, or $z_2\ind{i}\leq z_1\ind{i} +
x - \delta$. Furthermore, there is at least one index $i$
and one index $j$ such that $z_1\ind{i}=0$ and $z_2\ind{j}=0$. At
last, for $j \not\in V$, 
we have by construction $z_2\ind{j} \geq \min_{i\in V} z_2\ind{i}
-NM$, or $z_2\ind{j}=-\infty$. 
Alltogether, it shows that there are finitely many possible values
for $\p{z_2}=(z_2\ind{1},\dots, z_2\ind{n})$. 

Consequently, any acyclic path in $\Ua$ is of finite length. Since the
number of initial states is finite, it follows easily from K\"{o}nig
Lemma that the number of states of $\Ua$ is finite. 
\hspace*{\fill}\qed
\end{pf}

\begin{lem}\label{prop:unamb}
The automaton $\Ua$ is unambiguous.
\end{lem}

\begin{pf}
Define the surjective map
$$ \Psi \ :\begin{array}{ccl}\Ua & \longrightarrow & \cP \\
                       (\p{z},\p{p}) & \longmapsto & \p{p}\end{array}.$$
\\By construction of $\Ua$, the following properties hold. 

$i)$ The map $\Psi$ restricted to the initial states of $\Ua$ defines a
bijection between the initial states of $\Ua$ and $\cP$. 

$ii)$ Consider $\chemin{\p{p}\fleche{a}{.}\p{q}}{\cP}$. Then $\forall (\p{z},\p{p}) \in \Psi^{-1}(\p{p}), \exists ! (\p{z'},\p{q})\in
\Psi^{-1}(\p{q})$ such that $\chemin{(\p{z},\p{p})\fleche{a}{.}(\p{z'},\p{q})}{\Ua}$. 

$iii)$ A state $(\p{z},\p{q})$ is a final state of $\Ua$ if and only if $\p{q}$ is a final
state of $\cP$. 

\begin{figure}[h]
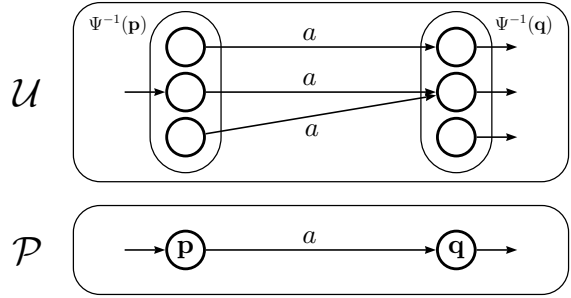

\begin{center}
\VCDraw{%
\begin{VCPicture}{(-1,-2)(7,5)}
\psframe[framearc=1](-.8,1.2)(.8,4.8)
\psframe[framearc=1](5.2,1.2)(6.8,4.8)
\psframe[framearc=.3](-2.5,1)(8.5,5)
\psframe[framearc=.6](-2.5,.5)(8.5,-1.5)
\State[\p{p}]{(0,-.5)}{R}
\State[\p{q}]{(6,-.5)}{S}
\State{(0,2)}{P1}
\State{(0,3)}{P2}
\State{(0,4)}{P3}
\State{(6,2)}{Q1}
\State{(6,3)}{Q2}
\State{(6,4)}{Q3}
\Initial{R}
\Final{S}
\EdgeL{R}{S}{a}
\EdgeR{P1}{Q2}{a}
\EdgeL{P2}{Q2}{a}
\EdgeL{P3}{Q3}{a}
\Initial{P2}
\Final{Q1}\Final{Q2}\Final{Q3}
\VCPut{(-1.5,4.5)}{\large$\Psi^{-1}(\p{p})$}
\VCPut{(7.5,4.5)}{\large$\Psi^{-1}(\p{q})$}
\VCPut{(-3.5,3)}{\Huge$\Ua$}
\VCPut{(-3.5,-.5)}{\Huge$\cP$}
\end{VCPicture}}
\end{center}
\caption{The properties of the map $\Psi$.}
\end{figure}

These three properties together imply that there is a bijection
between successful paths in $\P$ and successful paths in $\Ua$.
As $\P$ is unambiguous, so is $\Ua$.
\hspace*{\fill}\qed
\end{pf}

\begin{lem}\label{prop:sa}
 The automaton $\Ua$ recognizes the series $\bigoplus_{i\in I} S(\cA_i)$.
\end{lem}

\begin{pf}
Let $\ell$ be an integer and $u=a_0a_1\cdots a_{\ell-1}$ be a
word in the common support of the series $S(\cA_i)$.

By Lemma~\ref{prop:unamb}, there exists exactly one successful path labelled
by $u$ in the automaton $\Ua$:
    \[
    \pi=
    \chemin{
      \init{(\p{z_0},\p{q_0})}
      \flech{a_0} (\p{z_1},\p{q_1})\flech{a_1}\cdots
    \flech{a_{\ell-2}}(\p{z_{\ell-1}},\p{q_{\ell-1}})
      \flech{a_{\ell-1}} \final{(\p{z_\ell},\p{q_\ell})}}{\Ua}
    \]

$\bullet$ Fix $i\in\{1,\ldots ,n\}$.
Assume that $z_\ell\ind{i}=-\infty$.\\
Then $i$ is not a victorious
coordinate of $\pi$. Let $j$ be a victorious coordinate,
we show that $\coef{S(\A_i)}{u}<\coef{S(\A_j)}{u}$.
Hence the coefficient of $u$ in $\bigplus_{i\in I}S(\A_i)$ is not
realized by the coordinate $i$, which means that there is no damage in
having $z_\ell\ind{i}=-\oo$.

In the path $\pi$, there exists a minimal state $q_h$ such that the
coordinate $z\ind{i}_h$ is equal to $-\infty$. That means that the difference
between $z\ind{i}_h$ and $z\ind{j}_h$ would have been larger than $NM$.
Let $\pi'$ in $\P$ be the path that corresponds to $\pi$
(by the proof of Lemma~\ref{prop:unamb}, there is a canonical bijection
between successful paths of $\Ua$ and $\P$) and let $q'_h$ be
the state of $\pi'$ that corresponds to $q_h$. Let $\pi'_h$ be
the end of $\pi'$ from $q_h'$ onwards (including the final arrow).
Let us prove that the difference of weights on $\pi'_h$ between the
coordinates $i$ and $j$ is smaller than $NM$, that is:
\begin{equation}\label{eq:NM}
\wght{\pi'_h}\ind{i}-\wght{\pi'_h}\ind{j}\leqslant NM.
\end{equation} 
Actually, on every circuit, the weight with respect to $i$ is smaller
than or
equal to the weight with respect to $j$ (which is victorious),
and, if we delete all the circuits in $\pi'_h$, we obtain an acyclic
path that is necessarily shorter than $N-1$. On every transition,
the difference between the weights of the coordinates $i$ and $j$ is at most $M$.
Likewise, the difference between terminal functions is smaller than $M$.
Hence we proved~(\ref{eq:NM}). It means that the weight of coordinate
$i$ cannot catch up with the one of coordinate $j$. In particular, we have:
$\coef{S(\A_i)}{u}<\coef{S(\A_j)}{u}\leq \coef{\bigplus_{i\in I}
  S(\A_i)}{u}$.

$\bullet$ Assume that $z_\ell\ind{i}\neq-\infty$. Set
$\boldsymbol{\alpha}=(\alpha\ind{i}_{q_0\ind{i}})_{i\in I}$ and
$\boldsymbol{\beta}=(\beta\ind{i}_{q_\ell\ind{i}})_{i\in I}$. Let $\pi'$ be the
path in $\P$ that corresponds to $\pi$:
    \[
    \pi'=
    \chemin{
      \init[\boldsymbol{\alpha}]{\p{q_0}}
      \fleche{a_0}{\p{x_0}} q_1\fleche{a_1}{\p{x_1}}\cdots
      \fleche{a_{\ell-1}}{\p{x_{\ell-1}}}\final[\boldsymbol{\beta}]{\p{q_\ell}}}{\P}.
    \]

We have, by construction of the automaton $\Uc$:
\begin{align*}
     \coef{S(\A_i)}{u}=&\alpha_{q_0\ind{i}}\ind{i}+\sum_{k=0}^{\ell-1}x_k\ind{i}+\beta_{q_\ell\ind{i}}\ind{i}\\
                     =&\vmin{\boldsymbol{\alpha}}+z_0\ind{i}+\sum_{k=0}^{\ell-1}(y_k+z_{k+1}\ind{i}-z_k\ind{i})+\beta_{q_\ell\ind{i}}\ind{i}\\
                     =&\vmin{\boldsymbol{\alpha}}+\sum_{k=0}^{\ell-1}y_k+z_l\ind{i}+\beta_{q_\ell\ind{i}}\ind{i}
\end{align*}
Therefore, $\coef{S(\A_i)}{u}=\coef{\bigplus_{j\in I}S(\A_j)}{u}$ if
and only if $z_\ell\ind{i}+\beta_{q_\ell\ind{i}}\ind{i}=
\max_j[z_\ell\ind{j}+\beta_{q_\ell\ind{j}}\ind{j}]$. Now observe that
by construction, $$\coef{\Uc}{u} = \vmin{\boldsymbol{\alpha}} +
\sum_{k=0}^{\ell-1}y_k +\max_i
[z_\ell\ind{i}+\beta_{q_\ell\ind{i}}\ind{i}].$$ The equality
$\coef{\bigplus_{i\in I}S(\A_i)}{u}=\coef{\Uc}{u}$ follows easily.
\hspace*{\fill}\qed
\end{pf}

We now have all the ingredients to prove the proposition below. 

\begin{prop}\label{prop:wywiwyg}
Consider a finite family $(\A_i)_{i\in I}$ of trim and unambiguous
max-plus automata having the same support. Let $\P$ be the corresponding product
automaton. The series $\bigoplus_{i\in I}S(\A_i)$ is unambiguous if and only if
$\P$ satisfies the property ({\bf P}). In this case, the automaton $\Ua$ defined above is
finite, unambiguous, and realizes the series $\bigoplus_{i\in I}S(\A_i)$.
\end{prop}

\begin{pf}
Lemmas~\ref{prop:finite}, \ref{prop:unamb} and~\ref{prop:sa} show
that ({\bf P}) is a sufficient condition for 
$\bigoplus_{i\in I}S(\A_i)$ to be unambiguous. Let us prove that ({\bf
  P}) is also a necessary condition. 

By way of contradiction, assume that $S=\bigplus_{i\in I}S(\Ac_i)$ is
recognized by an unambiguous automaton $\Ua$ and that ({\bf P}) does
not hold.  There exists a path $\pi$ of $\P$
that can be decomposed into $\pi_0,\theta_1,\pi_1,\theta_2,\ldots ,\pi_r$,
where every $\theta_i$ is a circuit and $\bigcap_i \vict{\theta_i}=\emptyset$.
Let $u_i$ be the label of $\pi_i$ and $v_i$ the label of $\theta_i$. Let $s$ be the
maximal integer such that $V=\bigcap_{i\leqslant s}
\vict{\theta_i}\neq\emptyset$.
Let $w_{k,l}=u_0v_1^ku_1\cdots v_s^ku_sv_{s+1}^lu_{s+1}u_{s+2}\cdots
u_r$. For every $k,l$, $w_{k,l}$ is accepted by $\P$ and thus by $\Uc$
(with an unique successful path). Let $k_0,l_0$ be greater than the
number of states $d$ of $\Uc$. By the pigeon-hole principle, every
path in $\Uc$ labelled by $v_i^{k_0}$ (for $i\in\{1,\ldots ,s\}$) has a
sub-circuit labelled by $v_i^{k_i}$ (with $k_i<d$). Likewise, the path
labelled by $v_{s+1}^{l_0}$ has a sub-circuit labelled by
$v_{s+1}^{l_1}$. It means that there exist $(g_i,k_i,d_i)_{i\in[1,s]}$
and $(g_{s+1},l_1,d_{s+1})$ such that the successful path labelled by
$w_{k_0,l_0}$ in $\Uc$ has the following shape:
\begin{center}
\VCDraw{%
\begin{VCPicture}{(-1.5,-1)(19,3)}
\State{(0,0)}{A}
\State{(4,0)}{B}
\State{(8,0)}{C}
\State{(11,0)}{D}
\State{(18.5,0)}{E}
\Initial{A}
\Final{E}
\EdgeL{A}{B}{u_0v_1^{g_1}}
\EdgeL{B}{C}{v_1^{d_1} u_1v_2^{g_2}}
\EdgeL[.5]{D}{E}{v_{s+1}^{d_{s+1}} u_{s+1}u_{s+2}\cdots u_r}
\LoopN[.5]{B}{v_1^{k_1}}
\LoopN[.5]{C}{v_2^{k_2}}
\LoopN[.5]{D}{v_{s+1}^{l_1}}
\ChgEdgeLineStyle{dotted}
\EdgeL{C}{D}{}
\end{VCPicture}}
\end{center}

Let $K=\prod_{i\leq s}k_i$. Since $\Ua$ is unambiguous, for every pair
of integers $(\alpha,\beta)$, the word $w_{k_0+\alpha K,l_0+\beta l_1}$ is
accepted by a path that has the same shape; hence, there exist
$x=\coef{S}{w_{k_0,l_0}}$, $\rho$ and $\lambda$ such that, for every
$(\alpha,\beta)\in\N\times\N$, $\coef{S}{w_{k_0+\alpha K,l_0+\beta
    l_1}}=x+\alpha\rho+\beta\lambda$.

The word $w_{k_0+\alpha K,l_0+\beta l_1}$ labels in $\P$ a successful
path that is the concatenation of $\pi_0$, $(k_0+\alpha K)$ times $\theta_0$,
$\pi_1$,\ldots ,$\pi_s$, $(l_0+\beta l_1)$ times $\theta_{s+1}$,\ldots.
Therefore, for every $\beta$, there exists
$N_\beta$ such that, for every $\alpha>N_\beta$, the successful
coordinates of the path labelled by $w_{k_0+\alpha K,l_0+\beta l_1}$
belong to $V$ and the weight is equal to $y+\alpha\rho_1+\beta\lambda_1$,
where $y$ is a constant, $\rho_1$ is the sum of the
maximal weights of the circuits $\theta_1$ to $\theta_s$, and
$\lambda_1=\max_{i\in V}\wght{\theta_{s+i}}\ind{i}$.

Likewise, for every $\alpha$, there exists $M_\alpha$ such that, for
every $\beta>M_\alpha$, the successful coordinate of the path labelled
by $w_{k_0+\alpha K,l_0+\beta l_1}$ is a victorious coordinate of
$\theta_{s+1}$ and the weight of this path is equal to
$z+\alpha\rho_2+\beta\lambda_2$, where $z$ is a constant, $\rho_2$
is the maximum over the victorious coordinates of $\theta_{s+1}$ of
the sums of the weights of the circuits $\theta_1$ to $\theta_s$,
and $\lambda_2$ is the maximal weight of $\theta_{s+1}$.

To summerize, the following equalities hold:
\begin{align*}
\forall \alpha,\beta,\ \coef{S}{w_{k_0+\alpha K,l_0+\beta l_1}}&=x+\alpha\rho+\beta\lambda\\
\forall \beta,\forall\alpha>N_\beta,\ \coef{S}{w_{k_0+\alpha
    K,l_0+\beta l_1}}&=y+\alpha\rho_1+\beta\lambda_1\\
\forall \alpha,\forall\beta>M_\alpha,\ \coef{S}{w_{k_0+\alpha
    K,l_0+\beta l_1}}&=z+\alpha\rho_2+\beta\lambda_2
\end{align*}
Therefore, $\rho_1=\rho=\rho_2$ and
$\lambda_1=\lambda=\lambda_2$. Thus, there exists a coordinate that
belongs to $V$ and that is victorious on $\theta_{s+1}$; this
contradicts the maximality of $s$.

\medskip

It would be possible to use an argument similar to the one
in~\S\ref{sec:nR}, to prove the above.
\hspace*{\fill}\qed
\end{pf}

The main result is now a corollary of Proposition~\ref{prop:wywiwyg}:

\begin{thm}\label{th-carac}
  One can decide in an effective way, whether the series recognized by a
  finitely ambiguous max-plus automaton is unambiguous, and whether it
  is sequential.
\end{thm}

More precisely, turn first the finitely ambiguous automaton into an
equivalent finite union of unambiguous automata, all having the same support
(Corollary \ref{cor-famb2amb}).  Then check the property ({\bf P}) on the
new family of automata.  If ({\bf P}) is satisfied the series is unambiguous;
build the unambiguous automaton $\Ua$ (Proposition \ref{prop:wywiwyg}),
then decide the sequentiality of $\Ua$ (Theorem \ref{th-chof}).


\end{document}